\documentclass[lettersize,journal]{IEEEtran}
\usepackage{amsmath,amsfonts}
\usepackage{amsthm}
\usepackage{subcaption} 
\usepackage[hyphens]{url}       
\usepackage[ruled,norelsize,vlined,linesnumbered]{algorithm2e}
\makeatletter
\newcommand{\removelatexerror}{\let\@latex@error\@gobble}
\makeatother
\usepackage{amssymb}
\usepackage{array}
\usepackage{caption}
\usepackage{textcomp}
\usepackage{stfloats}
\usepackage{url}
\usepackage{verbatim}
\usepackage{graphicx}
\usepackage{cite}
\usepackage{multirow}
\usepackage{pifont}
\usepackage{makecell}

\usepackage[colorlinks,
            linkcolor=blue,
            anchorcolor=blue,
            citecolor=blue,
            urlcolor=black]{hyperref}

\makeatletter
\renewcommand{\eqref}[1]{%
    \textcolor{blue}{\hyperref[#1]{(\ref*{#1})}}%
}
\makeatother
\usepackage{booktabs} 
\usepackage{caption}  
\usepackage{tabularx}    
\usepackage{graphicx}
\usepackage[mathscr]{eucal}
\usepackage{tablefootnote}
\usepackage[labelformat=simple]{subcaption}

\begin{document}

\title{Generation Quality-Latency Tradeoff-Aware Inference Offloading for Multimodal LLMs in Cloud-Edge Continuum}

\author{Zhongxiao Wang, Yueshen Xu,~\IEEEmembership{Member,~IEEE,} Wei Xi,~\IEEEmembership{Member,~IEEE,} Xinkui Zhao, Tom H. Luan,~\IEEEmembership{Fellow,~IEEE}, Wei Shao,~\IEEEmembership{Member,~IEEE,} Rui Li,~\IEEEmembership{Member,~IEEE}
\IEEEcompsocitemizethanks{
\IEEEcompsocthanksitem This paper is funded by National Key Research and Development Program of China (2023YFF0905100), National Natural Science Foundation of China (62472338), Shaanxi Province Qinchuangyuan ``Scientist+Engineer'' Team Development Program (2024QCY-KXJ-165), and Open Foundation of Yunnan Key Laboratory of Software Engineering (2023SE301) (\textit{Corresponding author: Yueshen Xu, and Yueshen Xu contributes equally with Zhongxiao Wang, so he is also the co-first author}). 
\IEEEcompsocthanksitem Zhongxiao Wang, Yueshen Xu, and Rui Li are with the School of Computer Science and Technology, Xidian University, Xi'an 710126, China. E-mails: zhongxiaowang@stu.xidian.edu.cn, ysxu@xidian.edu.cn, and rli@xidian.edu.cn.
\IEEEcompsocthanksitem Wei Xi is with the School of Computer Science and Technology, Xi’an Jiaotong University, Xi’an 710049, China (e-mail: weixi.cs@gmail.com).
\IEEEcompsocthanksitem Xinkui Zhao is with the School of Software Technology, Zhejiang University, Ningbo 315048, China. E-mail: zhaoxinkui@zju.edu.cn.
\IEEEcompsocthanksitem Tom H. Luan is with the School of Cyber Science and Engineering, Xi'an Jiaotong University, Xi'an 710049, China. E-mail: tom.luan@xjtu.edu.cn.
\IEEEcompsocthanksitem Wei Shao is with the School of Computer Science and Engineering, University of New South Wales, Sydney 2052, Australia. E-mail: phdweishao@gmail.com.
}

}

\markboth{IEEE Transactions on Mobile Computing,~Vol.~XX, No.~XX, XX~2026}%
{Wang \MakeLowercase{\textit{et al.}}: Generation Quality-Latency Tradeoff-Aware Inference Offloading for Multimodal LLMs in Cloud-Edge Continuum}


\maketitle

\begin{abstract}
Beyond pure cloud, some efforts are being made to deploy Large Language Models (LLMs) in edge to accelerate inference response. So the deployment of LLMs in cloud-edge continuum becomes a promising paradigm, where the tasks involving multimodal data occupy a large part of requests. Under this continuum, users usually concern about multiple Quality-of-Service (QoS) attributes, but it is always intractable to jointly optimize them. In this paper, we propose to study the joint optimization of those attributes and focus on two key representatives, i.e., content generation quality and response latency. We propose to study the offloading technology to achieve a tradeoff between the two objectives in the cloud-edge collaborative Multimodal LLM (MLLM) system. However, it is highly difficult to predict generation quality and inference latency for MLLM inference tasks while optimizing this offloading process. To address these unprecedented difficulties, we propose a \textit{Quality-Latency Tradeoff-Aware MLLM Inference Offloading (QLMIO)} framework to make decisions that optimally balance generation quality and response latency. QLMIO consists of two novel capabilities: 1) \textit{MLLM Inference Latency Prediction (MILP)} and  2) \textit{MLLM Generation Quality Prediction (MGQP)}, which proactively predict inference latencies and generation correctness probabilities, respectively. Meanwhile, recognizing the absence of publicly available datasets tailored to the MLLM inference offloading problem, we constructed a real-world cloud-edge collaborative MLLM system and subsequently collected an \textit{MLLM Inference Offloading Benchmark (MIOBench)} to comprehensively evaluate our framework and facilitate the study of this problem. Extensive experimental results demonstrate that the QLMIO framework reduces latency by up to 58.14\% compared to baselines, while simultaneously matching the task completion rate achieved under the case that executes all requests exclusively on a cloud server. The dataset and codes are available at Github\footnote{https://anonymous.4open.science/r/MIOBench}.
\end{abstract}


\section{Introduction}
\IEEEPARstart{I}{n} recent years, with the unprecedented expansion of Large Language Models (LLMs)\cite{LLM_TPDS2026} around the world, the demand for LLM inference services is exponentially proliferating. In the current production environment, predominant LLM inference systems (e.g., Qwen\footnote{https://www.qianwen.com/} and ChatGPT\footnote{https://chatgpt.com/}) conventionally deploy hundred-billion-parameter LLMs purely on cloud servers to provide LLM inference services. However, this cloud-based deployment paradigm faces serval serious challenges, where two representatives
are as follows: 1) excessive response latency caused by overwhelming concurrent user requests within the same time slot, and 2) unsustainable resource and cost overheads arising from provisioning numerous hundred-billion-parameter LLM instances. Meanwhile, edge computing has emerged as a distributed computing paradigm that can partially offload computation workloads from cloud centers to edge nodes, thereby reducing response latency and resource costs \cite{Cloudedgellm_TIT2025}. This raises a promising solution, that is, deploying LLMs with heterogeneous parameter scales across the cloud-edge continuum to collaboratively deliver LLM inference services \cite{Cloudedgellm_icpads2025}.

As LLM inference tasks are becoming increasingly multimodal, Multimodal Large Language Models (MLLMs) \cite{MLLM_TVT2026} have emerged as a preferred choice for users owing to their advanced cross-modal processing capabilities. In practical applications of cloud-edge collaborative MLLM systems, due to the significant computational capacity disparity between cloud and edge servers, the parameter scales of deployable 
MLLMs exhibit substantial heterogeneity. For instance, a cloud server equipped with an RTX5090 GPU featuring 32GB VRAM can effortlessly deploy the 30-billion-parameter Qwen3-VL\footnote{https://huggingface.co/collections/Qwen/qwen3-vl, VL is short for Vision-Language} model, while an edge device with merely 8GB VRAM (e.g., Jetson Orin Nano Super\footnote{https://www.nvidia.com/en-us/autonomous-machines/embedded-systems/jetson-orin/nano-super-developer-kit/}) can only deploy the 2-billion-parameter Qwen3-VL model. This heterogeneity in device computational capabilities and model parameter scales inevitably leads to significant variations in generation quality and response latency among different nodes within cloud-edge collaborative MLLM systems. 

\begin{figure}[htbp]
\centering
\begin{subfigure}{0.48\textwidth}
    \includegraphics[width=\linewidth]{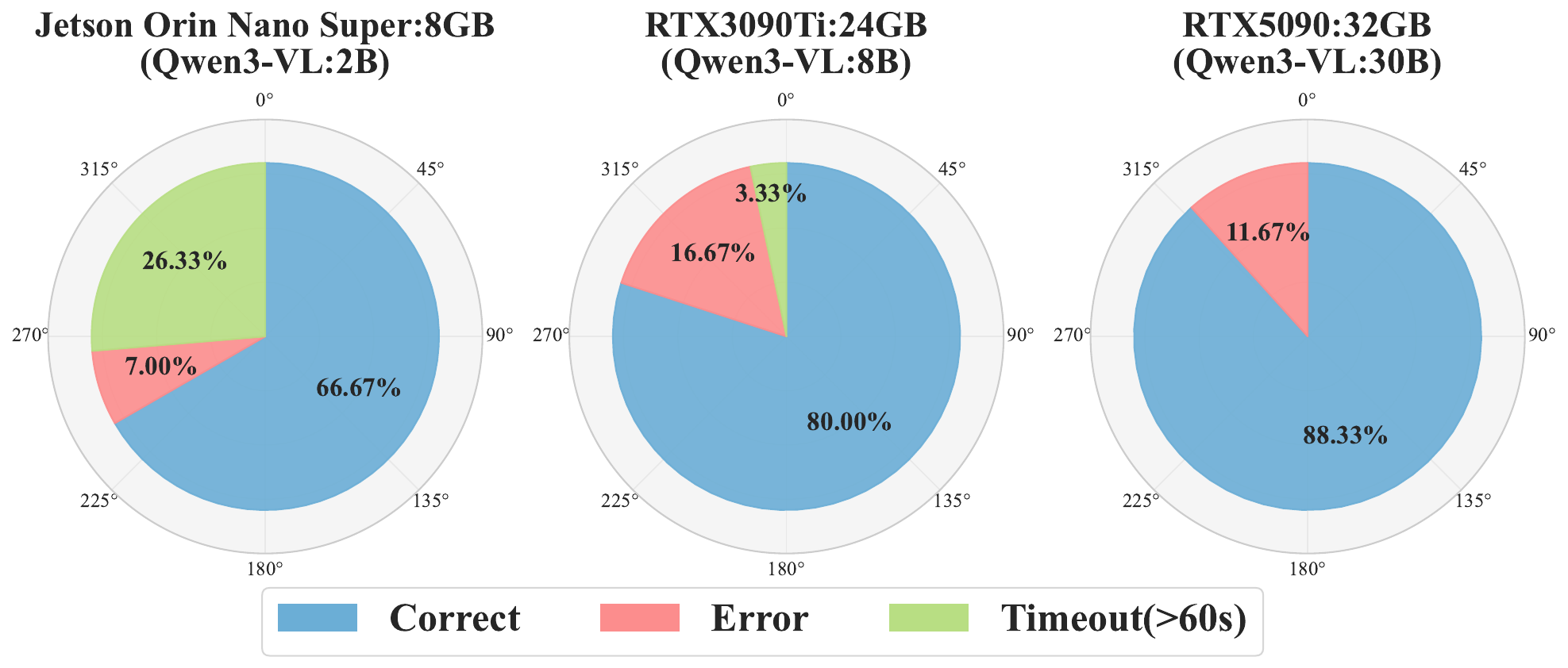}
    \caption{Generation quality of different devices}
    \label{QoSDis:sub1}
\end{subfigure}

\vspace{0.5cm}

\hspace{-0.30cm} 
\begin{subfigure}{0.48\textwidth}
    \includegraphics[width=\linewidth]{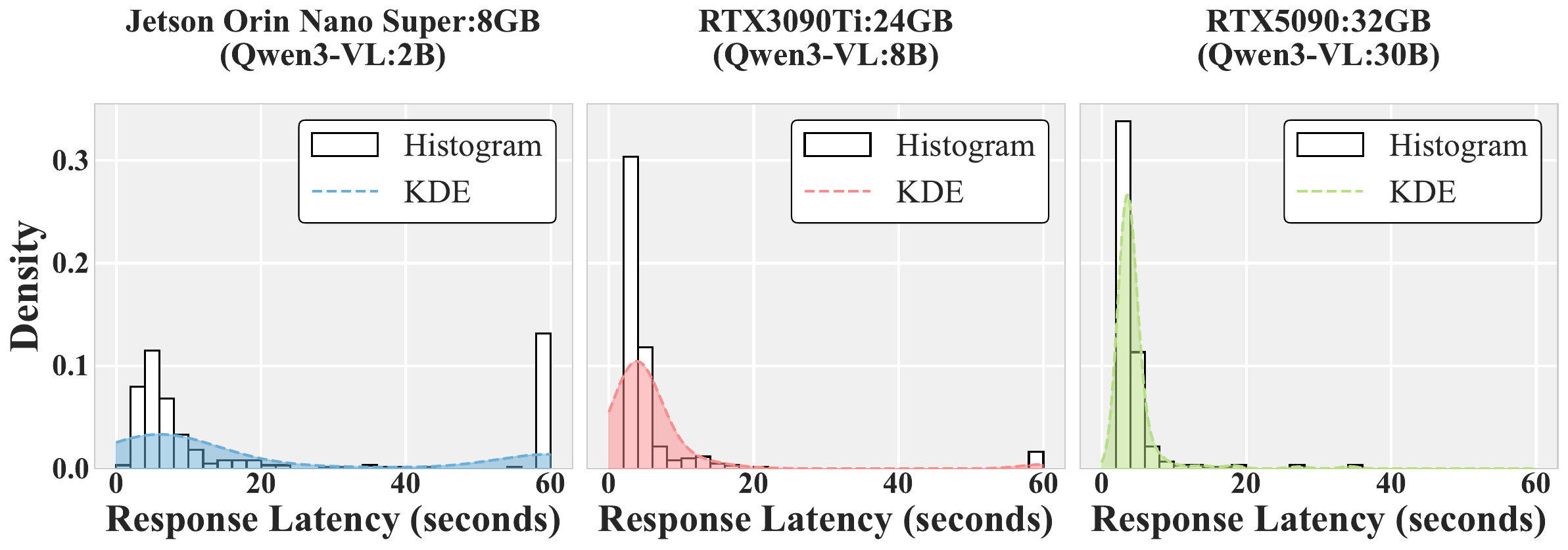}
    \caption{Response latency of different devices}
    \label{QoSDis:sub2}
\end{subfigure}
\caption{Performance differences of generation quality and response latency among different devices in a cloud-edge collaborative MLLM system.}
\label{QoSDis}
\end{figure}

To visually demonstrate performance disparities among heterogeneous nodes, we randomly sampled 300 inference tasks from the MMBench\footnote{MMBench is a multimodal dataset containing 3,377 distinct choice questions composed of both images and text. A detailed description of MMBench is provided in Section \ref{Sec:Experiment}.} dataset \cite{mmbench} and executed them on three distinct devices (RTX5090: 32GB, RTX3090Ti: 24GB, and Jetson Orin Nano Super: 8GB) hosting Qwen3-VL variants with disparate parameter scales (30B, 8B, and 2B). The generation quality and response latency results are shown in Fig. \ref{QoSDis}. As observed in Fig. \ref{QoSDis:sub1}, on low-computing devices such as the Jetson Orin Nano Super, the generation accuracy of the MLLM is only 66.67\%, with a high proportion of time-out samples reaching 26.33\%. In contrast, on the high-performance device RTX5090, the generation accuracy approaches 90\%, and no time-out samples occur. This phenomenon clearly demonstrates a significant disparity in generation accuracy when the same task is performed on different devices equipped with different MLLMs. Fig. \ref{QoSDis:sub2} displays the histograms and the Kernel Density Estimation (KDE) plots of response latency across different devices. 

We can find that nearly all inference tasks on the RTX5090 exhibit response latencies below 10s with high kurtosis, indicating stable and consistent inference performance. In contrast, the Jetson Orin Nano Super shows a dispersed latency distribution, characterized by low response efficiency and significant latency fluctuations. The above phenomena are attributable to two critical factors: \textbf{1) the cognitive capabilities of MLLMs vary significantly with their parameter sizes}, leading to substantial differences in generation quality across different MLLMs, and \textbf{2) a lightweight MLLM deployed on an edge device requires an extended Chain-of-Thought to process sophisticated tasks}, leading to longer inference latency\cite{COT_ICLR2025}. This substantial heterogeneity across nodes makes the offloading of MLLM inference an imperative technology for the efficient operation of cloud-edge collaborative MLLM systems.

Compared to traditional edge computing offloading problems \cite{Offloading_TPDS2025}\cite{Offloading_TNSE2025}, the offloading for MLLM inference presents two unprecedented challenges: \textbf{1) there are no existing methods can predict the generation quality of a specific task on one given node}, making it exceptionally difficult to guarantee the overall generation accuracy in cloud-edge collaborative MLLM systems, and  \textbf{2) the response latency of MLLM inference cannot be simply derived from the data volume and the devices' computational capacities}, which introduces significant uncertainty into the response time of cloud-edge collaborative MLLM systems.  To address these challenges and ensure low-latency, high-reliability inference in cloud-edge collaborative MLLM systems, this paper makes the following primary contributions:

\begin{itemize}
    \item[$\bullet$] \textbf{To the best of our knowledge, we are one of the pioneers to address MLLM inference offloading for cloud-edge continuum, considering both generation quality and response latency.} We formulate this problem as a multi-objective optimization problem and propose a Quality-Latency Tradeoff-Aware MLLM Inference Offloading (QLMIO) framework to efficiently address this problem, thereby guaranteeing the system's performance.
\end{itemize}
\begin{itemize}
    \item[$\bullet$] \textbf{To address the challenge that the generation quality and inference latency are hard to predict, caused by heterogeneous cognitive capabilities across nodes, we design an MLLM Generation Quality Prediction (MGQP) module and an MLLM Inference Latency Prediction (MILP) module.}  MGQP and MILP can estimate task-node compatibility by extracting semantic features from user input prompts and node-level features, enabling the prediction of generation quality and inference latency for specific tasks on designated nodes.
\end{itemize}
\begin{itemize}
    \item[$\bullet$] \textbf{Given the current absence of a public dataset that contains the actual generation quality and response latency of various MLLM inference tasks across nodes with diverse configurations, we collected and created an MLLM Inference Offloading Benchmark (MIOBench).} MIOBench is collected from a real-world cloud-edge collaborative MLLM system constructed by this paper using a variety of devices and MLLMs.
\end{itemize}
\begin{itemize}
    \item[$\bullet$] \textbf{Given the high costs of deploying MLLMs in cloud-edge environments, we developed a Cloud-Edge Collaborative MLLM System simulator (CEMLLM-Sim) based on our MIOBench dataset for training and evaluating MLLM inference offloading algorithms.}  We open-sourced the MIOBench dataset, the CEMLLM-Sim simulation framework, and all source codes on GitHub\footnote{https://anonymous.4open.science/r/MIOBench}. 
\end{itemize}

The subsequent sections are arranged as follows. Section \ref{Sec:RelatedWork} provides an overview of the related work. Section \ref{Sec:SystemModelandProblemFormulation} presents the system model and formulates the optimization problem. Section \ref{Sec:Framework} provides a comprehensive description of the proposed QLMIO framework. Section \ref{Sec:Experiment} demonstrates the complete experimental configuration and results analysis. Section \ref{Sec:Conclusion} concludes the article and discusses future work.

\section{Related Work}
\label{Sec:RelatedWork}
In this section, we review the literature relevant to our problem from two perspectives: (i) LLM inference in cloud and edge environments, and (ii) offloading for LLM inference in cloud and edge environments.

\subsection{LLM Inference in Cloud and Edge Environments}
To address issues such as long response latencies in LLM systems deployed on cloud servers, some researchers have proposed leveraging edge computing to enhance the system's performance. Cai et al. \cite{EDGELLM_ICWS2024} designed a server-node collaboration framework for LLM serving to efficiently utilize edge resources to accelerate LLM fine-tuning and inference in resource-constrained scenarios. Jin et al.\cite{EDGELLM_ICWS2025} proposed a cloud-edge collaboration framework for LLMs, which uses a latency-aware early exit mechanism and a cloud context management approach to reduce communication overhead and preserve LLM generation accuracy. Zhang et al. \cite{EDGELLM_IOTJ2025} proposed an LLM inference framework for edge computing, which partitions a computation-intensive LLM into affordable shards and deploys them on edge devices. Zhang et al. \cite{EDGELLM_TWC2025} proposed an LLM edge inference framework, which incorporates batching and model quantization to ensure high-throughput inference on resource-limited edge devices. Zheng et al. \cite{EDGELLM_TNSE2026} proposed a collaborative cloud-edge adaptation framework for LLMs for mobile networking, which reduces LLM adaptation costs while maintaining task performance in mobile network scenarios through cloud-edge collaboration. Although these methods attempt to integrate some advantages of edge computing with LLMs, they have not yet studied the offloading task or explored the offloading technology, so they leave a large room to achieve a more optimized performance of cloud-edge collaborative LLM systems.

\subsection{Offloading for LLM Inference in Cloud and Edge Environments}

Computation offloading has been widely adopted in edge networks or cloud-edge collaborative systems to enhance the Quality-of-Service (QoS) \cite{Offloading0_TMC2025}\cite{Offloading1_TMC2025}. In recent years, some researchers have also attempted to apply the offloading technology to cloud-edge collaborative LLM systems\cite{yang2026msaoadaptivemodalitysparsityaware}. Wang et al. \cite{llmoffload_VTC2024} proposed an optimization approach for task offloading and diffusion model generation, which tried to balance multiple metrics such as latency and resource consumption in edge computing. He et al. \cite{llmoffload_TMC2024} developed an inference framework to a cloud-edge collaborative LLM system, and tried to optimize task offloading and resource allocation decisions. Yang et al.\cite{llmoffload0_TMC2025} proposed a Deep Reinforcement Learning (DRL)-based QoS-aware LLM routing framework aimed at handling response of LLM services while maintaining content generation. Xu et al. \cite{llmoffload1_TMC2026} proposed a collaborative inference framework for LLMs and small language models in mobile edge computing scenarios, which jointly decides where to cache models and how to offload inference tasks. Besides offloading, a few researchers also explore some other related tasks for LLM inference. Huang et al. \cite{llmoffload_iotj2025} proposed a two-timescale optimization framework for LLM inference in edge–cloud networks, where they primarily focus on task scheduling and resource allocation. While these related works have attempted to use computation offloading to enhance the QoS in LLM systems, none of them investigates how to simultaneously guarantee both response latency and the quality of generated content in a real-world cloud-edge collaborative MLLM system. Therefore, this paper fills this gap.

\section{System Model and Problem Formulation}
\label{Sec:SystemModelandProblemFormulation}    

\begin{figure}[htbp] 
    \centering
    \includegraphics[width=0.45\textwidth]{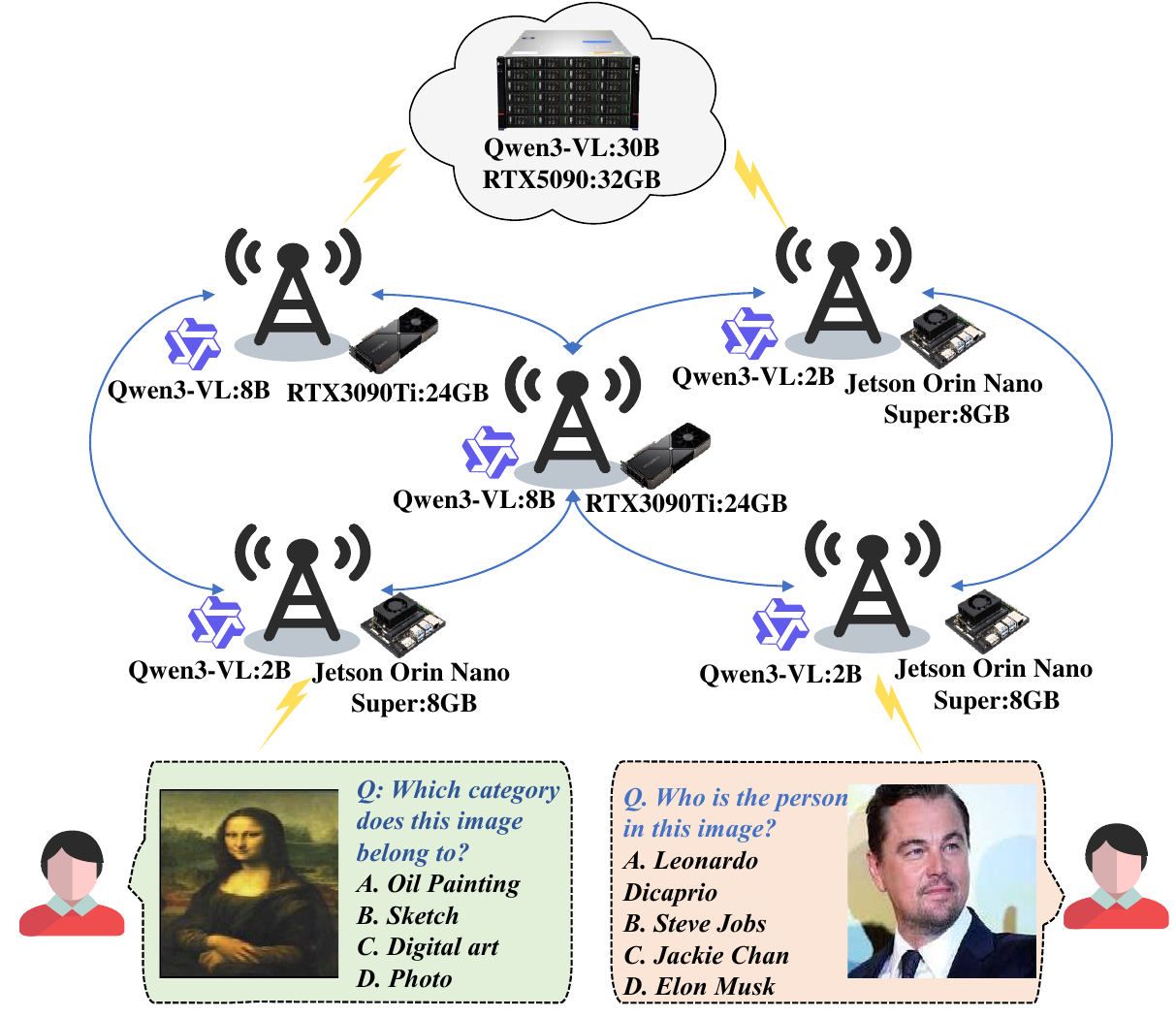}
    \caption{A toy example of the real-world cloud-edge collaborative MLLM system.} 
    \label{fig:cloudedge} 
\end{figure}

We give a toy example for the real-world cloud-edge collaborative MLLM system, which is shown in Fig. \ref{fig:cloudedge}, where MLLM inference tasks proposed by users can be offloaded to edge servers or a cloud server. Note that the illustrated images in Fig. \ref{fig:cloudedge} are from the MMBench dataset.

Specifically, $C$ is used to represent the cloud server in this system, $E=\{e_1,e_2,...,e_\mathcal{E}\}$  denotes the set of edge servers,  $M=\{m_1, m_2,...,m_\mathcal{M}\}$  defines the set of distinct MLLMs,  $D=\{d_1, d_2,...,d_\mathcal{D}\}$  represents the set of device types, and $U=\{1,2,...,\mathcal{U}\}$  represents the set of users. For each server, it is represented by a 3-tuple. For example, an edge server $e_i=\left(m_{e_i}, d_{e_i}, q_{e_i}\right)$ indicates that the server $e_i$ is equipped with a device $d_{e_i}\in D$ and is deployed with an MLLM $m_{e_i} \in M$, and the inference tasks that have been offloaded to this server form a set $q_{e_i}$. In this system, each of $\mathcal{U}$ users can propose an MLLM inference request at anytime. We define the request from user $u$ as a 2-tuple $R_{u}=\left(P_{u}, I_{u}\right)$, where $P_{u}$ and $I_{u}$ denote the text and image in the requests, respectively. When we offload an MLLM inference task $R_{u}$ to an idle server $x_{R_u}\in\{E,C\}$, the response latency $T^r_{x_{R_u}}$ can be calculated as:
\begin{equation}
    \label{responselayentcy}
    T^r_{x_{R_u}} = T^c_{x_{R_u}}+T^t_{x_{R_u}}
\end{equation}
where $T^c_{x_{R_u}}$ and $T^t_{x_{R_u}}$ are the computation/inference latency and the transmission latency, respectively, when the task $R_u$ is offloaded to the server $x_{R_u}$. When the server $x_{R_u}$ is not idle, the total offloading latency $T_{x_{R_u}}$ can be calculated as:
\begin{equation}\label{offloading}
    T_{x_{R_u}} = T^r_{x_{R_u}}+T^q_{x_{R_u}}
\end{equation}
where $T^q_{x_{R_u}}$ is the queuing latency, which can be calculated as:
\begin{equation}\label{queue}
    T^q_{x_{R_u}}=\sum\limits_{R'_u \in q_{x_{R_u}}}T^r_{x_{R'_u}}
\end{equation}
where $q_{x_{R_u}}$ is the task queue in the server $x_{R_u}$. Then we use a binary variable $b_{x_{R_u}} \in \{0,1\}$ to quantify the quality of the content generated by the MLLM deployed on server $x_{R_u}$ when processing task $R_u$. A value of $b_{x_{R_u}}=1$ indicates that the MLLM generates the correct content, while  $b_{x_{R_u}}=0$ denotes that the content generated by the MLLM is incorrect or overtime.

Based on the aforementioned system model, we formulate a multi-objective optimization problem $\mathbf{P}$, aiming to minimize the average task offloading latency and the average task failure rate. The problem $\mathbf{P}$ is defined as follows:
{\begin{alignat}{2}
\mathbf{P}: 
\quad & \min_{X}
\begin{cases} 
\frac{1}{\mathcal{U}}\sum\limits_{u\in U}T_{x_{R_u}}, \\[10pt] 
\frac{1}{\mathcal{U}}\sum\limits_{u\in U}-b_{x_{R_u}}.
\end{cases} & \label{obj} \\
\mbox{s.t.}\quad
& x_{R_u} \in \{E,C\} & \label{st1}
\end{alignat}}where $X$ is the set of all offloading decisions. The objective function \eqref{obj} is designed to minimize the average task offloading latency and the average task failure rate, where $\min\limits_{X}\frac{1}{\mathcal{U}}\sum\limits_{u\in U}-b_{x_{R_u}}$ aims to maximize the number of ones among all $b_{x_{R_u}}$. The constraint \eqref{st1} is an integer constraint that guarantees each task can only be offloaded to one execution server. As observed in Problem $\mathbf{P}$, the two optimization objectives exhibit inherent conflicts. For instance, to minimize the task failure rate, the most straightforward approach is to offload all tasks to the cloud server equipped with the most powerful MLLM to guarantee the correctness of the generated content. However, this would lead to a surge in queuing latency, thereby dramatically increasing the total offloading latency. Such a conflict underscores the urgent need for an offloading solution capable of achieving a tradeoff between offloading latency and generation quality.

While extensive research has proposed some approaches to address multi-objective optimization problems for edge computing offloading \cite{Multiobject_TVT2025}\cite{Multiobject1_TVT2025}, the problem $\mathbf{P}$ studied in this paper differs significantly from the traditional edge computing offloading problems majorly owing to the following reason: Due to the varying cognitive capacities of different MLLMs and the diverse semantics of inference tasks, the offloading latency $T_{x_{R_u}}$ and the quality of content generated by MLLMs $b_{x_{R_u}}$ in the objective function are difficult to quantify directly. To address this complex multi-objective optimization problem, this paper: 1) collects an MLLM Inference Offloading Benchmark (MIOBench), and 2) designs an MLLM Generation Quality Prediction (MGQP) module and an MLLM Inference Response Latency Prediction (MILP) module, and 3) proposes a Quality-Latency Tradeoff-Aware MLLM Inference Offloading (QLMIO) framework.

\section{Design for the Developed Framework}\label{Sec:Framework}
\subsection{The Designed MGQP Module and MILP Module}

\begin{figure}[!th]
  \centering
  \includegraphics[width=0.99\columnwidth]{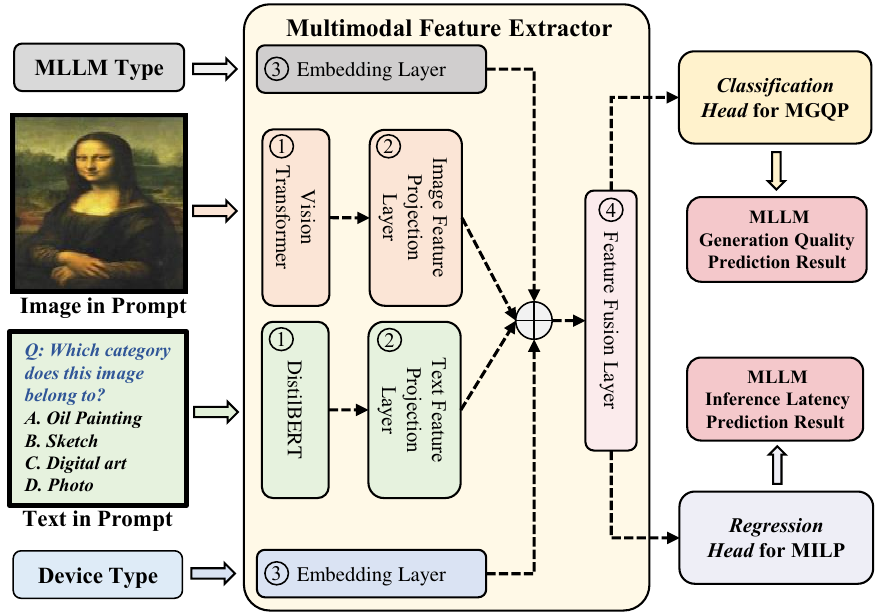}
  \caption{Core neural network architectures of MGQP and MILP}
  \label{fig:PredictionModule}
\end{figure}

To address the challenges posed by the difficulty of predicting the generation quality and inference latency of MLLMs, we design an MGQP module and an MILP module. Their core neural network architecture is illustrated in Fig. \ref{fig:PredictionModule}. 

Specifically, the inputs of both the MGQP and MILP modules can be represented as a 4-tuple $g_{x_{R_u}}=\left(I_u, P_u,m_{x_{R_u}},d_{x_{R_u}}\right)$, where $I_u \in \mathbb{R}^{H\times W \times 3}$ denotes the image in the prompt of user $u$, $P_u$ represents the textual prompt, $m_{x_{R_u}}$ and $d_{x_{R_u}}$ indicate the MLLM type and device type on server $x_{R_u}$, respectively. We have two objectives in this paper: one is to leverage the designed MGQP module to predict the quality of MLLM generation, and is denoted as $\hat{b}_{x_{R_u}} \in \{0,1\}$ corresponding to the failure (timeout or incorrect) category and the success category, respectively; and the other is to employ the MILP module to predict the MLLM inference response time $\hat{T}^r_{x_{R_u}}$. As illustrated in Fig. \ref{fig:PredictionModule}, the universal multimodal feature extractor of MGQP and MILP employs a multi-branch (four branches here) fusion architecture that integrates visual, textual, and meta-information (i.e., MLLM type and device type) features through a fusion pipeline. This universal multimodal feature extractor comprises four key components: 

\textbf{\ding{172} Frozen pre-trained encoders for modality-specific feature extraction:} For textual feature extraction, we employ a frozen DistilBERT\cite{Sanh2019DistilBERTAD} (i.e., distilbert-base-uncased\footnote{https://huggingface.co/distilbert/distilbert-base-uncased}) as our textual backbone to encode the text $P_u$ in the input prompt. DistilBERT is a lightweight model with rapid inference, making it suitable for real-time computational offloading scenarios. Given the tokenized input text $\mathcal{P}_u=(\mathcal{P}^1_u,\mathcal{P}^2_u,...,\mathcal{P}^L_u)$ with the attention mask $\mathcal{Q}_u=(\mathcal{Q}^1_u,\mathcal{Q}^2_u,...,\mathcal{Q}^L_u)$, we can use the DistilBERT model to obtain a contextualized token representation $\mathcal{H}_{P_u}\in \mathbb{R}^{L\times768}$ through:
\begin{equation}\label{distilbert}
    \mathcal{H}_{P_u}=\text{DistilBERT}(\mathcal{P}_u,\mathcal{Q}_u)
\end{equation}
where 768 is the default setting in the original paper of DistilBERT, and $L$ denotes the maximum length of the input text. Subsequently, the mean pooling in Eq. \eqref{meanpooling} is applied to $\mathcal{H}_{text}$ to derive a fixed-length text embedding $f_{P_u}\in\mathbb{R}^{768}$.
\begin{equation}
\label{meanpooling}
    f_{P_u} = \frac{\sum ^{L}_{i
    =1}\mathcal{Q}_i\cdot\mathcal{H}^{(i)}_{P_u}}{\sum ^{L}_{i
    =1}\mathcal{Q}_i}
\end{equation}
For visual feature extraction, we utilize a frozen Vision Transformer\cite{VIT} (i.e., vit-base-patch16-224\footnote{https://huggingface.co/google/vit-base-patch16-224}) as our visual backbone to encode the image $I_u$ in the input prompt. The image is first divided into a set of non-overlapping patches, each of which is linearly projected to a 768-dimensional embedding. Then we use the Vision Transformer model to extract the [CLS] token representation $f_{I_u} \in \mathbb{R}^{768}$ as the global image feature through:
 \begin{equation}
     f_{I_u} = \text{ViT}(I_u)_\text{[CLS]}
 \end{equation}
 
\textbf{\ding{173} Learnable feature projection modules:} 
To align features from different modalities into a common latent space, we incorporate a learnable feature projection module following both the Vision Transformer and DistilBERT encoders. Each modality-specific feature (i.e., text feature $f_{P_u}$ and image feature $f_{I_u}$) is projected through a fully-connected layer with LayerNorm, Gaussian Error Linear Unit (GELU) activation, and Dropout regularization. This process can be mathematically formulated as follows:

\begin{equation}
\begin{aligned}
    f'_{P_u} = 
    \text{Dropout}\Big(\text{GELU}\big(\text{LayerNorm}(
    W_{t_1} f_{P_u} + b_{t_1}\big) \big)\Big)
\end{aligned}
\end{equation}
and
\begin{equation}\label{projection}
\begin{aligned}
    f'_{I_u} = 
    & \text{Dropout}\Big( \text{GELU}\big(\text{LayerNorm}(
    W_{i_1} f_{I_u} + b_{i_1}\big) \big)\Big)
\end{aligned}
\end{equation}
where $f'_{P_u},f'_{I_u}\in \mathbb{R}^{64}$ denote the projected features, and $W_{t_1}, W_{i_1}\in \mathbb{R}^{64\times768}$ are the projection weight matrices. This reduction in dimensionality serves as an information bottleneck that filters irrelevant features while preserving discriminative signals.

\textbf{\ding{174} Categorical embedding layers for meta-information encoding:}
To incorporate meta-information about the inference environment, we encode the MLLM type $m_{x_{R_u}}$ and device type $d_{x_{R_u}}$ into $m'_{x_{R_u}}\in \mathbb{R}^{32}$ and $d'_{x_{R_u}}\in \mathbb{R}^{32}$ through two learnable embedding layers, respectively. This process can be formulated as follows:
\begin{equation}\label{embed1}
    m'_{x_{R_u}} = \text{Embedding}_{model}(m_{x_{R_u}})
\end{equation}
and
\begin{equation}\label{embed2}
    d'_{x_{R_u}} = \text{Embedding}_{device}(d_{x_{R_u}})
\end{equation}
The categorical features are concatenated to form $f'_{{x_{R_u}}}=[m'_{x_{R_u}};d'_{x_{R_u}}]\in \mathbb{R}^{64}$, which represents systematic variations across different inference environment configurations.

\textbf{\ding{175} A feature fusion layer for cross-modal integration:}
The projected features from all branches are concatenated and projected to a fused feature representation $f^{fused}_{x_{R_u}}\in \mathbb{R}^{64}$ through a fusion network. This process can be formulated as follows:
\begin{equation}\label{fused}
    f^{fused}_{x_{R_u}} = \text{Fusion}([f'_{P_u},f'_{I_u},f'_{x_{R_u}}])
\end{equation}
The fusion module $\text{Fusion}(\cdot)$ consists of two linear layers with LayerNorm, GELU activation, and Dropout regularization. This design enables effective cross-modal interaction while mitigating overfitting through aggressive regularization.

Following the aforementioned neural network architecture, we can construct the MGQP and MILP modules using specific output heads. The details are as follows:

\textbf{(1) The classification head for the MGQP module}:
For the task of MLLM generation quality prediction, we incorporate a classification head that maps the fused feature representation $f^{fused}_{x_{R_u}}$ to a discrete quality label $\hat{b}_{x_{R_u}}\in\{0,1\}$. Specifically, a two-layer (i.e., $W^\text{log}_1$ and $W^\text{log}_2$) fully-connected layer with GELU activation projects $f^{fused}_{x_{R_u}}$ into logits $f^\text{log}_{x_{R_u}}\in \mathbb{R}^2$ over 2 quality categories (i.e., failure and success). This process can be formulated as follows:
\begin{equation}
    f^\text{log}_{x_{R_u}} = W^\text{log}_2\cdot\text{Dropout}\Big(\text{GELU}(
    W^\text{log}_{1} f^{fused}_{x_{R_u}}+ b^\text{log}_1\big) \Big) + b^\text{log}_2 
\end{equation}
where $W^\text{log}_{1}\in \mathbb{R}^{32\times64}$ and $W^\text{log}_{2}\in \mathbb{R}^{2\times32}$ are the weight matrices. During the training phase of the MGQP module, we use Focal Loss \cite{FocalLoss_TMC2026}. Focal Loss is extensively employed to mitigate the issue of class imbalance in classification tasks, with its weighting mechanism giving greater weight to examples from classes with small sample sizes. This loss function is defined as:
\begin{equation}
    \mathcal{L}_\text{Focal}(g_{x_{R_u}}) = -\alpha(1-p_{g_{x_{R_u}}})^{\gamma}\text{log}(p_{g_{x_{R_u}}})
\end{equation}
where $\alpha$ serves as a weight in the class proportion to balance the influence of different classes during training, $p_{g_{x_{R_u}}}$ is the model's estimated probability for the ``success'' class, and $\gamma$ is a hyperparameter to tune the sensitivity of the loss function to misclassified examples.

\textbf{(2) The regression head for the MILP module}: For the task of MLLM inference latency prediction, we incorporate a regression head that maps the fused feature representation $f^{fused}_{x_{R_u}}$ to a continuous latency label $\hat{T}^r_{x_{R_u}}$. Specifically, a two-layer fully-connected layer with GELU activation projects $f^{fused}_{x_{R_u}}$ into a latency value $\hat{T}^r_{x_{R_u}} \in \mathbb{R}$. This process can be mathematically formulated as follows:
\begin{equation}
    \hat{T}^r_{x_{R_u}} = W^t_2\cdot\text{GELU}(
    W^t_{1} f^{fused}_{x_{R_u}}+ b^t_1\big) + b^t_2 
\end{equation}
where $W^\text{t}_{1}\in \mathbb{R}^{32\times64}$ and $W^\text{t}_{2}\in \mathbb{R}^{1\times32}$ are the weight matrices. During the training phase of the MGQP module, we use Huber Loss\cite{HuberLoss_TMC2025}. The Huber loss function serves as the cornerstone for robust regression training, effectively balancing the advantages of Mean Squared Error (MSE)\cite{MSE_TSC2025} and Mean Absolute Error (MAE)\cite{MAE_TPDS2025} to mitigate the influence of outliers (i.e.,  timeout samples where ${T}^r_{x_{R_u}}>60 \text{s}$). Formally, the Huber loss is defined as:
\begin{equation}\label{Huber}
    \mathcal{L}_\text{Huber}(g_{x_{R_u}}) = \begin{cases} 
\frac{1}{2}\psi^2, & \text{if } |\psi| \leq \delta, \\
\delta|\psi| - \frac{1}{2}\delta^2, & \text{otherwise}.
\end{cases}
\end{equation}
where $\psi = {T}^r_{x_{R_u}} - \hat{T}^r_{x_{R_u}}$ represents the residual between the predicted latency $\hat{T}^r_{x_{R_u}}$ and the ground truth ${T}^r_{x_{R_u}}$, and $\delta$ is a hyperparameter dictating the transition point between MSE loss and MAE loss.

\subsection{The Proposed Quality-Latency Tradeoff-Aware MLLM Inference Offloading (QLMIO) Framework}

To address the complex multi-objective optimization problem $\mathbf{P}$, we propose a QLMIO framework based on DRL. We use a 3-tuple $\left(\mathcal{S}, \mathcal{A}, \mathcal{R}\right)$ to represent the Markov Decision Process (MDP) of the MLLM inference offloading, where $\mathcal{S}=\{s_{t}\}_{t\in T}$ is the state space, $\mathcal{A}=\{a_{t}\}_{t\in T}$ is the action space, $\mathcal{R}=\{r_{t}\}_{t\in T}$ is the reward set, and $T$ is the set of agent decision time slots. The detailed definitions are as follows:
\begin{figure*}[!t]
  \centering
  \includegraphics[width=0.89\textwidth]{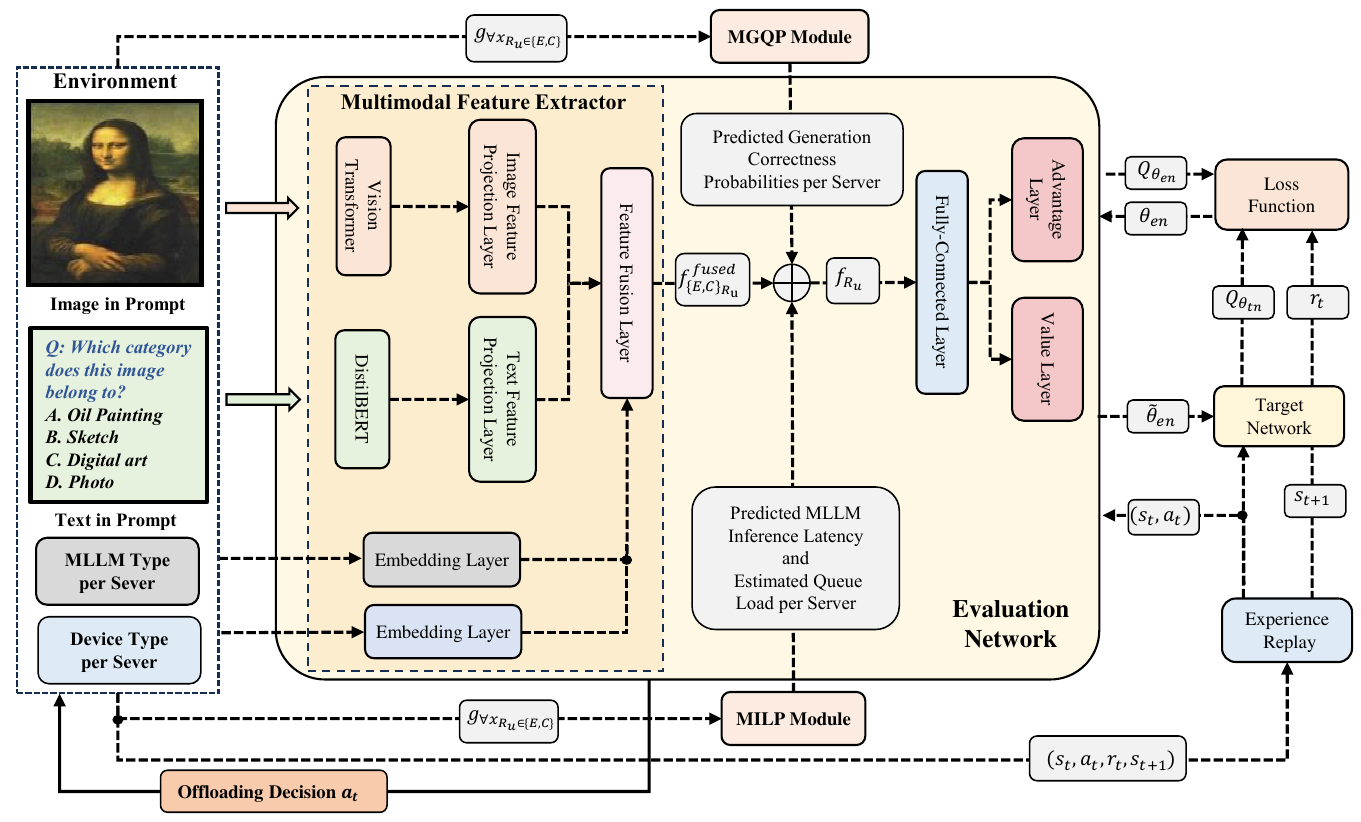}
  \caption{Developed architecture of the proposed QLMIO framework}
  \label{fig:QLMIO}
\end{figure*}

\textbf{1) State $s_{t}$:}
To comprehensively assess the completion quality of an MLLM inference task within the cloud-edge collaborative MLLM system, the state of the QLMIO in time slot $t$ is defined as follows:
\begin{equation}\label{state}
\begin{split}
    s_{t} = \big( &P_u, I_u, m_{\forall x_{R_u}\in \{E,C\}}, d_{\forall x_{R_u}\in \{E,C\}}, \\
    &\hat{T}^{r}_{\forall x_{R_u}\in \{E,C\}},\tilde{T}^{q}_{\forall x_{R_u}\in \{E,C\}}, \hat{b}_{\forall x_{R_u}\in \{E,C\}} \big)
\end{split}
\end{equation}
where $P_u$ and $I_u$ represent the text and image in the user prompt, respectively;  $m_{\forall x_{R_u}\in \{E,C\}}$ and $ d_{\forall x_{R_u}\in \{E,C\}}$ denote the  MLLM type and the device type per server, respectively; $\hat{T}^{r}_{\forall x_{R_u}\in \{E,C\}}$ is the predicted MLLM inference latency per server using our designed MILP module; $\tilde{T}^{q}_{\forall x_{R_u}\in \{E,C\}}$ is the estimated queue load per server. The estimated queue load $\tilde{T}^{q}_{x_{R_u}}$ for one server $x_{R_u}$ is defined as the average predicted latency per queued task:
\begin{equation}\label{queueload}
    \tilde{T}^{q}_{x_{R_u}} = \hat{T}^{q}_{x_{R_u}}/|q_{x_{R_u}}|
\end{equation}
where $\hat{T}^{q}_{x_{R_u}}$ is the queue latency estimated by Eq. \eqref{queue} and $|q_{x_{R_u}}|$ is the length of the task queue; $\hat{b}_{\forall x_{R_u}\in \{E,C\}}$ in the state denotes the generation correctness probabilities for the current task on each server predicted by our designed MGQP module. 

\textbf{2) Action $a_{t}$:}
The proposed QLMIO framework determines the optimal server in $E$ or $C$ for MLLM inference offloading according to state information $s_t$. Therefore, we can formulate the offloading decision in time slot $t$ as the following discrete action:
\begin{equation}
    a_t\in \{E,C\}
\end{equation}

\textbf{3) Reward $r_{t}$:}
Considering that the objective of problem $\mathbf{P}$ is to minimize
the average task offloading latency and the average task failure
rate, while the goal of the DRL agent is to maximize the long-term reward, we formulate the immediate reward $r_t$ for performing the offloading action $a_t$ on task  $R_{u}$ as follows:
\begin{equation}\label{reward}
    r_t = 1- T_{{a_t}_{R_u}}/T_{\text{Greedy}(R_u)} + r^b_t
\end{equation}
where $T_{{a_t}_{R_u}}$ denotes the actual total offloading latency incurred after task $R_{u}$ is offloaded to server $a_t$, which is calculated by Eq. \eqref{offloading}; $\text{Greedy}(\cdot)$ represents a greedy algorithm that offloads each task to the server with the shortest task queue, providing a total offloading latency $T_{\text{Greedy}(\cdot)}$ (according to Eq. \eqref{offloading}) for comparison; $r^b_t$ is the task completion reward, which equals $+1$ upon successful task completion and $-2$ in case of task failure, thereby incentivizing reliable inference delivery. Specifically, the primary rationale for setting the failure penalty to $-2$ is to maintain a balanced trade-off between the latency reward component (i.e., $1 - T_{{a_t}_{R_u}} / T_{\text{Greedy}(\cdot)}$) and the task completion reward. During the early exploration stages of training, the DRL agent may yield a decision latency $T_{{a_t}_{R_u}}$ that is several times larger than the greedy baseline $T_{\text{Greedy}(\cdot)}$, which severely amplifies the negative magnitude of the latency reward. Assigning a substantial penalty of $-2$ ensures that the requirement for task completion is not overwhelmed by extreme latency penalties, guiding the agent to jointly optimize both objectives.

Next, we need to determine an optimal offloading action in each decision time slot. Since the state space of the proposed QLMIO framework is continuous and the offloading action is discrete, we choose the Dueling Double Deep Q-Network (D3QN) \cite{D3QN_TMC2025} as the backbone. The core neural network architecture of QLMIO is shown in Fig. \ref{fig:QLMIO}. 
Fig. \ref{fig:QLMIO} shows: \textbf{1)} The evaluation network of QLMIO framework employs the same multimodal feature extractor as the extractor contained in the MILP and MGQP modules to process text $P_u$ and image $I_u$ in the state $s_t$ (according to Eq.\eqref{distilbert} to Eq.\eqref{projection}). \textbf{2)} Furthermore, QLMIO encodes the device types $d_{\forall x_{R_u}\in \{E,C\}}\in \mathbb{R}^{(\mathcal{E}+1)\cdot 16}$ and model types $m_{\forall x_{R_u}\in \{E,C\}}\in \mathbb{R}^{(\mathcal{E}+1)\cdot 16}$ of all servers in state $s_t$ within the cloud-edge collaborative MLLM system through two embedding layers (according to Eq.\eqref{embed1} and Eq. \eqref{embed2}). $+1$ represents the cloud. \textbf{3)} After that, the features from all branches are concatenated and projected to a fused feature representation $f^{fused}_{\{E,C\}_{R_u}}\in \mathbb{R}^{32}$ (according to Eq. \eqref{fused}), allowing unified representation of heterogeneous system components.
\textbf{4)} Subsequently, the fused feature representation $f^{fused}_{\{E,C\}_{R_u}}$ is concatenated with both the inference latency $\hat{T}^{r}_{\forall x_{R_u}\in \{E,C\}}\in \mathbb{R}^{\mathcal{E}+1}$ predicted by MILP for task $R_u$ offloaded to each server and the queue load $\tilde{T}^{q}_{\forall x_{R_u}\in \{E,C\}}\in \mathbb{R}^{\mathcal{E}+1}$ estimated by Eq. \eqref{queueload} for each server, along with the generation quality  $\hat{b}_{\forall x_{R_u}\in \{E,C\}}\in \mathbb{R}^{\mathcal{E}+1}$ predicted by MGQP for task $R_u$ offloaded to each server, forming a combined representation $f_{R_u}\in \mathbb{R}^{32+3\cdot(\mathcal{E}+1)}$ that is then fed into a fully-connected layer dedicated to feature extraction. \textbf{5)} This fully-connected layer consists of two linear layers with LayerNorm, GELU activation, and Dropout regularization. The parameter matrices of the two layers correspond to the dimensions of $256 \times (32 + (\mathcal{E}+1)\cdot3)$ and $256 \times256$, respectively. \textbf{6)} Finally,  QLMIO splits the network architecture into two branches to calculate the value function and the advantage function to learn both the state value and nuanced discrepancies between different actions.

\begin{figure}[h]
    \removelatexerror
    \begin{algorithm}[H]
        \SetAlgoLined 
        \small
        \caption{The training process of the developed QLMIO framework}\label{DRL}
        \LinesNumbered
        \KwIn{Set of users $U$, number of episodes $\mathbb{E}$, set of edge servers $E
        $, cloud server $C$, training interval $\mathbb{S}$, learning rate $\eta$ and batch size $\mathbb{B}$}
        \KwOut{Parameters of the evaluation network $\theta_{en}$}
        Initialize a DRL model with an experience replay $\mathbb{M}$, and the trained MILP and MGQP modules\; 
        $step \gets 0$\;
        \For{$episode=1:\mathbb{E}$}{
            $T^q \gets [0,\dots ,0] \in \mathbb{R}^{\mathcal{E}+1}$\;
            $ T_{\text{Greedy}(TQ_U)}\gets[0,\dots,0]\in \mathbb{R}^{\mathcal{U}}$\;
            $t\gets0,s_t \gets \varnothing, TQ_U \gets \varnothing$\;
            Each of $\mathcal{U}$ users independently proposes an MLLM inference task within the environment and a task queue $TQ_{U}$ is constructed\;
            \For{$u=1:\mathcal{U}$}{
                $R_u\gets TQ_{U}\left[u\right]$\;
                Obtain $T_{\text{Greedy}(R_u)}$ according to Eq. \eqref{reward}\;
                $T_{\text{Greedy}(TQ_U)}[u] \gets T_{\text{Greedy}(R_u)}$\;
            }
            \For{$u = 1:\mathcal{U}$}{
                $R_u\gets TQ_{U}\left[u\right]$\;
                Generate state $s_t$ according to Eq. \eqref{state}\;
                Select an action $a^*_t$ according to Eq. \eqref{action} and offload the task to server $a^*_t$\;
                Retrieve the actual response latency $T^r_{{a^*_t}_{R_u}}$ and actual generation quality $b_{{a^*_t}_{R_u}}$ from the ground truth data (MIOBench dataset)\;
                $T_{{a^*_t}_{R_u}} \gets T^r_{{a^*_t}_{R_u}} + T^q[a^*_t]$\;
                $T^q[a^*_t] \gets T^r_{{a^*_t}_{R_u}} + T^q[a^*_t]$\;
                Obtain $r^b_t$ in Eq. \eqref{reward}\;
                $r_t \gets 1-T_{{a^*_t}_{R_u}}/T_{\text{Greedy}(TQ_U)}[u] + r^b_t$\;
                Observe state $s_{t+1}$ from environment\;
                Store $\left(s_t,a^*_t,r_t,s_{t+1}\right) $ in $\mathbb{M}$\;
                
                \If{$|\mathbb{M}|>\mathbb{B}$ \& $step \equiv 0 \pmod{\mathbb{S}}$}{
                    Sample $\mathbb{B}$ transitions from $\mathbb{M}$\;
                    Update $\theta_{en}$ according to Eq. \eqref{Huber}\;
                }
                $s_t \gets s_{t+1}, t \gets t+1, step \gets step+1$\;
            }
            $\theta_{tn} \gets \tau \cdot \theta_{en} + (1-\tau) \cdot \theta_{tn}$\;
            //$\tau$ \textit{is a hyperparameter used for soft update}
        }
    \end{algorithm}
\end{figure}

Now, we can present the complete training process for the proposed QLMIO framework in Algorithm \ref{DRL}. The DRL model of QLMIO is initialized (line 1) with its architecture depicted in Fig. \ref{fig:QLMIO}. Within each episode, the QLMIO framework initializes list $T^q$ (initialized to zeros and its length is equal to $\mathcal{E}+1$) to record actual queuing latency for each server (line 4), and a list $T_{\text{Greedy}(TQ_U)}$ (initialized to zeros and its length is equal to $\mathcal{U}$) to record the total offloading latency for generating the greedy algorithm decision for each task (line 5).  Subsequently, each of $\mathcal{U}$ users independently proposes an MLLM inference task, and all tasks are aggregated into a collective task queue $TQ_U$ (lines 6-7). The QLMIO framework then offloads tasks using the greedy algorithm and populates $T_{\text{Greedy}(TQ_U)}$ with the resulting latencies (lines 8-12) for comparison. After that, for each task in $TQ_U$ with state $s_t$, the QLMIO framework computes the value function $V_{\theta_{en}}(s_t)$ and the advantage function $A_{\theta_{en}}(s_t, a^j_t)$ using the value layer and advantage layer, respectively, where $\theta_{en}$ represents the parameters of the evaluation network. Combining the value function $V_{\theta_{en}}(s_t)$ and the advantage function $A_{\theta_{en}}(s_t, a^j_t)$ enables the Q function $Q_{\theta_{en}}(s_t, a^j_t)$ to be calculated as follows:

\begin{equation}
    Q_{\theta_{en}}(s_t, a^j_t) = V_{\theta_{en}}(s_t) + A_{\theta_{en}}(s_t, a^j_t) + \frac{1}{\mathcal{E}+1}\sum_{i=1}^{\mathcal{E}+1}A_{\theta_{en}}(s_t, a^i_{t})
\end{equation}
where $a^j_t$ represents the action of offloading the task to the server $a^j_t \in \{E,C\}$. Then, the QLMIO framework selects an action $a^*_t$ based on the $\epsilon-greedy$ strategy \cite{egreedy_TPDS2024} (lines 14-16).  This action selection strategy is characterized as follows:
\begin{equation}\label{action}
    a^*_t = 
    \begin{cases} 
    \mathop{\arg\max}\limits_{a^j_t} Q_{\theta}(s_t, a^j_t), & \text{with probability } 1 - \epsilon \\
    \text{random action}, & \text{with probability } \epsilon
    \end{cases}
\end{equation}

Since the MIOBench dataset we collected contains the actual response latency and generation quality resulting from offloading user tasks to a server in a real-world cloud-edge collaborative MLLM system, we can directly retrieve the actual response latency $T^r_{{a^*_t}_{R_u}}$ and generation quality $b_{{a^*_t}_{R_u}}$ from the MIOBench dataset (line 17). After that, the total offloading latency $T_{{a^*_t}_{R_u}}$ and task completion reward $r^b_t$ can be calculated, from which the reward $r_t$ for executing action $a^*_t$ is derived (lines 18-21). Then, the QLMIO framework observes the next state $s_{t+1}$ and stores the transition $\left(s_t,a^*_t,r_t,s_{t+1}\right) $ in the experience replay $\mathbb{M}$ (lines 22-23). When the size of experience replay $\mathbb{M}$ exceeds the batch size $\mathbb{B}$ and the number of decision steps exactly divides the training interval
$\mathbb{S}$, the QLMIO framework samples a batch of transitions from $\mathbb{M}$ (lines 24-25). Subsequently, all parameters of the evaluation network are updated through backpropagation to minimize the Huber loss function (line 26). Finally, the parameters of the evaluation network are transferred to the target network using soft update \cite{softupdate_TNSE2025} at the end of each episode (line 30).

\section{Benchmark and Experiment}
\label{Sec:Experiment}
\subsection{An MLLM Inference Offloading Benchmark}
To assist the QLMIO framework in making optimal offloading decisions, it is imperative to train predictors for both MLLM inference latency and generation quality (i.e., the proposed MILP and MGQP modules). However, there are no public datasets tailored for the MLLM inference offloading problem. To fill this gap, we constructed a comprehensive MLLM inference offloading benchmark, \textbf{MIOBench}.

We selected the MMBench\cite{mmbench} dataset as the user task set. MMBench is a comprehensive multimodal dataset comprising 3,377 unique choice questions that integrate visual and textual elements. It encompasses 20 diverse MLLM inference task categories such as action recognition, future prediction, and spatial relation reasoning. Its broad scope effectively empowers a comprehensive and multi-dimensional evaluation of QLMIO. Furthermore, we established a real-world cloud-edge collaborative MLLM system, with detailed configurations provided in Table \ref{table:PARAMETER}. Architecturally, edge nodes are directly interconnected through the same-band Wi-Fi Local Area Network (LAN), while edge-to-cloud communication is provisioned over the cross-domain Wi-Fi Wide Area Network (WAN). After that, we deployed various MLLMs across servers of this cloud-edge system using Ollama\footnote{https://ollama.com/} according to Table \ref{table:PARAMETER}. 
\begin{table}[htbp]
  \centering
  \caption{Server configurations}
  \label{table:PARAMETER}
  \resizebox{\linewidth}{!}{
    \begin{tabularx}{\linewidth}{c>{\centering\arraybackslash}X>{\centering\arraybackslash}X}
      \toprule
      \multicolumn{1}{c}{} & \multicolumn{1}{c}{\textbf{Device}} & \multicolumn{1}{c}{\textbf{MLLM}} \\
      \midrule
      \multirow{2}{*}{Edge} 
      & Jetson Orin Nano Super: 8GB & Qwen3-VL:2B \\
      & RTX 3090Ti: 24GB & Qwen3-VL:8B \\
      \midrule
      \multirow{1}{*}{Cloud} 
      & RTX 5090Ti: 32GB & Qwen3-VL:30B \\

      \bottomrule
    \end{tabularx}%
  }
\end{table}

Afterward, each task from MMBench is offloaded to every server to obtain the actual inference latency $T^r_{x_{R_u}}$ and generation quality $b_{x_{R_u}}$. These offloading traces constitute the MIOBench dataset. Specifically, each sample in MIOBench is formalized as a key-value pair, and the field descriptions are illustrated in Table \ref{tab:MIOBench_fields}.
\begin{table}[htbp]
    \centering
    \caption{Description of fields in our MIOBench dataset}
    \label{tab:MIOBench_fields}
    \begin{tabularx}{\columnwidth}{@{}>{\raggedright\arraybackslash}p{1.8cm}>{\centering\arraybackslash}p{1.5cm}X@{}}
        \toprule
        \textbf{Key} & \textbf{Type} & \textbf{Description} \\
        \midrule
        dataset      & String  & Source dataset name (i.e., MMBench). \\
        prompt       & String  & Complete model input including context and options. \\
        device\_type & String  & Hardware deployed for inference (e.g., RTX3090ti). \\
        model\_name  & String  & Name and scale of the evaluated MLLM (e.g., Qwen3-VL:8B). \\
        score      & Integer & generation quality (1: success, 0: failure, -1: timeout). \\
        latency\_ms  & Float   & Inference response latency in milliseconds. \\
        sample\_id   & Integer & Unique identifier for the specific sample. \\
        index        & Integer & Sequential row index within the dataset. \\
        source       & String  & Original task source (e.g., YouTube\footnotemark). \\
        \bottomrule
    \end{tabularx}
\end{table}
\footnotetext{https://news.youtube.com}

Ultimately, we compiled \textbf{10,131} MLLM inference offloading records from 3,377 diverse tasks, which constitute the MIOBench dataset. This dataset can be accessed from GitHub\footnote{https://anonymous.4open.science/r/MIOBench}.

\subsection{Developed CEMLLM-Sim and Experiment Settings}
Since the actual inference response latency and generation quality resulting from offloading each task to any server in our cloud-edge collaborative MLLM system are already recorded in MIOBench, we developed a Cloud-Edge Collaborative MLLM System simulator (\textbf{CEMLLM-Sim}), dedicated to training and evaluating MLLM offloading algorithms, and to facilitating the research on the related research in the community. Within this simulator, we obtain the exact latency and quality metrics for any offloading decision by retrieving MIOBench, thereby avoiding the exorbitant costs associated with building a large-scale cloud-edge collaborative system. In all experiments presented in this paper, MIOBench was split into training, validation, and test sets at a ratio of 8:1:1 to thoroughly evaluate algorithm performance. Additionally, CEMLLM-Sim supports user-defined system configurations. We train and evaluate our algorithms primarily in three different cloud-edge MLLM system configurations, as summarized in Table \ref{tab:system_config}.
\begin{table}[htbp]
    \centering
    \caption{System configurations}
    \label{tab:system_config}
    \renewcommand{\arraystretch}{1}
    \begin{tabular}{@{}lllccc@{}}
        \toprule
        \multirow{2}{*}{\textbf{Device}} & \multirow{2}{*}{\textbf{MLLM}} & \multirow{2}{*}{\textbf{Role}} & \multicolumn{3}{c}{\textbf{Server Count}} \\ 
        \cmidrule(lr){4-6}
        & & & \textbf{5} & \textbf{10} & \textbf{15} \\ 
        \midrule
        RTX 5090 & Qwen3-VL:30B & Cloud & 1 & 1 & 1 \\
        RTX 3090Ti & Qwen3-VL:8B & Edge & 1 & 2 & 4 \\
        Jetson Orin Nano Super & Qwen3-VL:2B & Edge & 3 & 7 & 10 \\
        \bottomrule
    \end{tabular}
\end{table}

Subsequently, the proposed MILP and MGQP modules are trained and evaluated on the constructed MIOBench, while the proposed QLMIO framework is trained and evaluated on the CEMLLM-Sim simulator. All experiments were implemented using PyTorch 2.9.0 with CUDA 13.0, an Intel Xeon Silver 4510 @4.10 GHz CPU, 128 GB of RAM, and an NVIDIA RTX Pro 6000 GPU. The representative hyperparameter settings are shown in Table \ref{tab:hyperparameter}.
\begin{table}[htbp]
    \centering
    \caption{Representative hyperparameter settings}
    \label{tab:hyperparameter}
    \begin{tabularx}{\columnwidth}{@{} l l X @{}}
        \toprule
        \textbf{Parameter} & \textbf{Value} & \textbf{Description} \\
        \midrule
        $L$ & 256 & The maximum length of the input text \\
        $\delta$     & 1.0    & A hyperparameter in the Huber loss \\
        $\epsilon$   & 1.0    & A hyperparameter in the $\epsilon$-greedy \\
        $\eta$   & 0.0001    & Learning rate of QLMIO \\
        $\mathbb{S}$ & 5      & Training interval in the training process of QLMIO \\
        $\mathbb{E}$ & 12000      & Number of episodes of QLMIO \\
        $\mathbb{B}$ & 256      & Batch size in the training process of QLMIO \\
        $|\mathbb{M}|$ &  10000    & Size of the replay buffer in QLMIO \\
        $\tau$       & 0.005  & A hyperparameter used for soft update \\
        \bottomrule
    \end{tabularx}
\end{table}

\subsection{Baselines}
For performance comparison, we compare the proposed QLMIO framework with the following baselines:

\textbf{1) QoS-Aware LLM Routing Algorithm based on DRL (QoS-Aware RL) \cite{llmoffload0_TMC2025}}: QoS-Aware RL is one of the state-of-the-art algorithms for LLM routing in pure edge computing environments. It employs graph attention to extract the QoS features of ordinary LLMs and uses DRL to perform LLM routing for text-only queries in edge environments.

\textbf{2) Dueling Double Deep Q-Network (D3QN) \cite{D3QN_TPDS2026}}: D3QN is one of the most effective value-based DRL algorithms. It mitigates the Q-value overestimation through double DQN and accelerates convergence in dynamic cloud-edge environments by employing a dueling architecture.

\textbf{3) Soft Actor-Critic (SAC) \cite{SAC_TSC2025}}: SAC is one of the most advanced off-policy DRL algorithms. It simultaneously optimizes policy entropy maximization and cumulative reward objectives through a decoupled actor-critic architecture.

\textbf{4) All-Could \cite{Cloud_TNSE2026}}: All cloud is a traditional method that offloads all  MLLM inference tasks to the central cloud servers, entirely neglecting the computational capabilities in edge networks. 

\textbf{5) Greedy}: We constructed the Greedy algorithm to compare the performance of our proposed QLMIO framework with traditional heuristic methods. The Greedy algorithm employs a greedy strategy to offload tasks to the server with the shortest task queue.

\subsection{Training Results of MILP and MGQP Modules}

\begin{figure}[!htb]
    \centering
    \subfloat[Training and validation loss]{%
        \includegraphics[width=0.47\columnwidth]{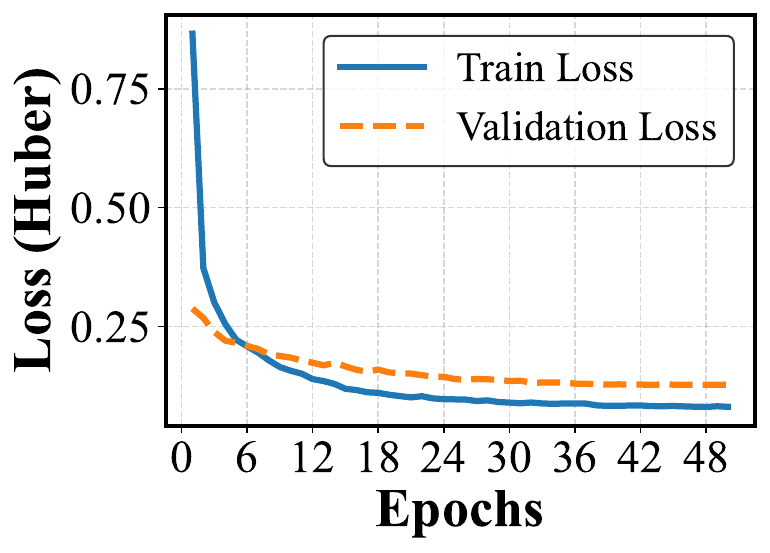}%
        \label{fig:milp_loss}%
    }
    \hfil
    \subfloat[Training and validation MAE]{%
        \includegraphics[width=0.47\columnwidth]{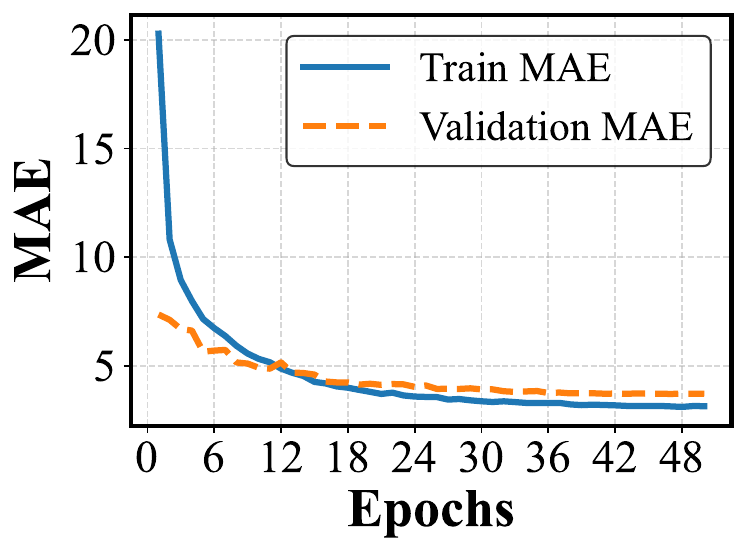}%
        \label{fig:milp_mae}%
    }
    \caption{Training convergence of the proposed MILP Module.}
    \label{fig:milp_training}
\end{figure}

We illustrate the training convergence of our proposed MILP module over 50 epochs in Fig.~\ref{fig:milp_training}. The MILP module exhibits rapid and robust stabilization, where both the Huber Loss and MAE show sharp drops within the initial 15 epochs without severe overfitting degradation. Notably, the validation MAE converges in a substantial 3.70. Given the heavy-tailed nature of the MLLM inference latency distribution (cf. Fig. \ref{QoSDis:sub2}), which frequently varies in the tens of seconds depending on the complexity of the input prompt, this latency prediction module exhibits high availability. Consequently, this reliable latency predictor can act as a robust foundation and assist the QLMIO framework to make latency-aware offloading decisions in the cloud-edge collaborative MLLM system.

\begin{figure}[!htb]
    \centering
    \subfloat[Training and validation loss]{%
        \includegraphics[width=0.47\columnwidth]{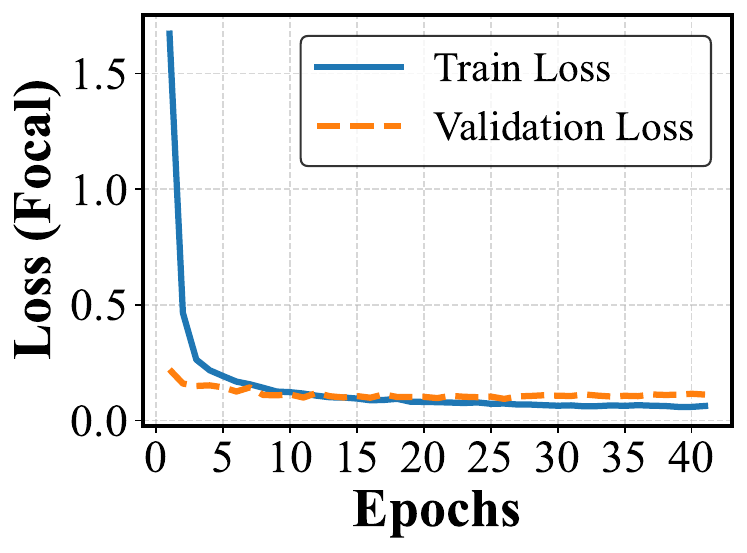}%
        \label{fig:milp_loss}%
    }
    \hfil
    \subfloat[Training and validation accuracy]{%
        \includegraphics[width=0.47\columnwidth]{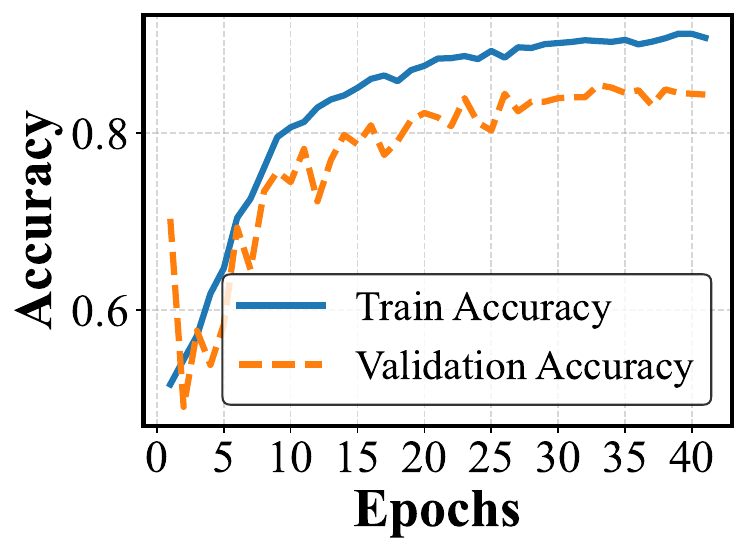}%
        \label{fig:milp_mae}%
    }
    \caption{Training convergence of the proposed MGQP Module.}
    \label{fig:MGQP_training}
\end{figure}

Furthermore, we illustrate the training convergence of proposed MGQP module in Fig.~\ref{fig:MGQP_training}. The MGQP module exhibits rapid convergence within the initial 30 epochs, effectively minimizing its Focal Loss while attaining a peak classification accuracy of 85.46\%. Consequently, this quality predictor can empower the subsequent QLMIO framework to make quality-aware offloading decisions by proactively estimating the potential probability of success before actual offloading.

\subsection{Convergence Analysis of the QLMIO Framework}

\begin{figure}[!htb]
    \centering
    \subfloat[Average reward]{%
        \includegraphics[width=0.47\columnwidth]{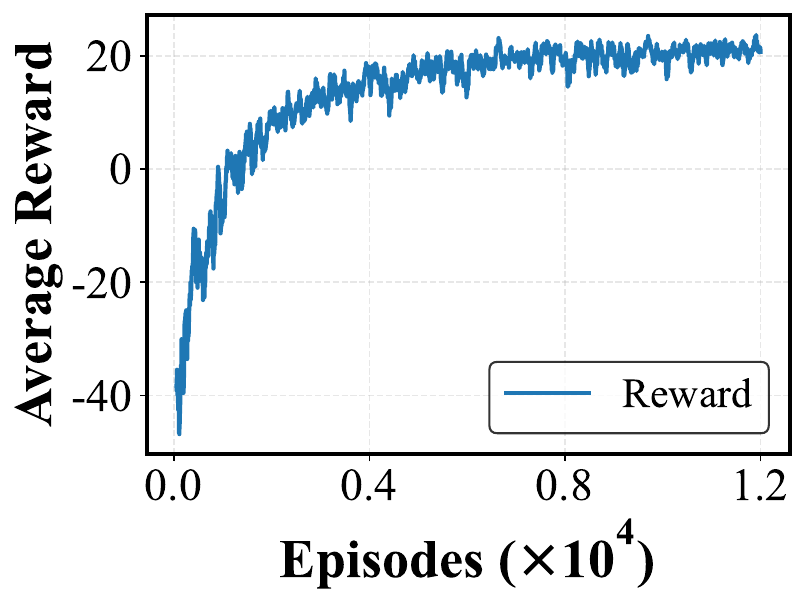}%
        \label{fig:rlreward}%
    }
    \hfil
        \subfloat[Average loss ]{%
        \includegraphics[width=0.47\columnwidth]{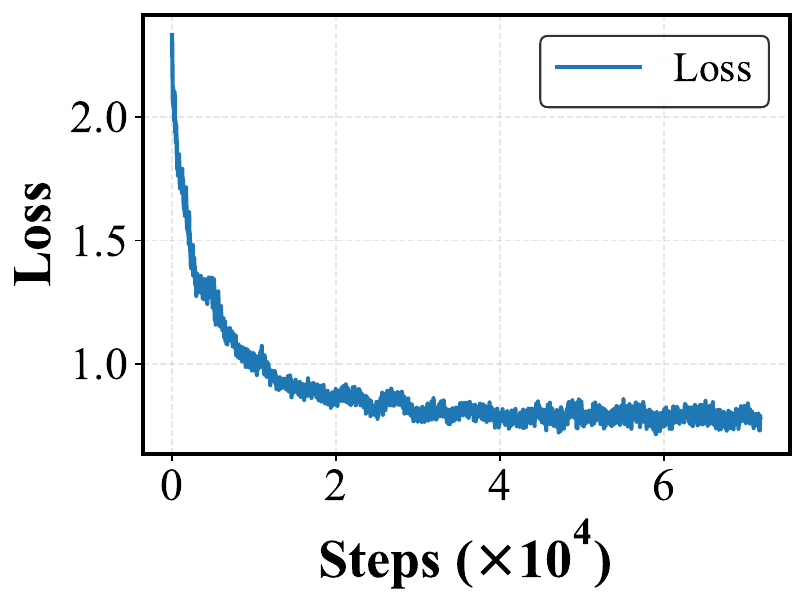}%
        \label{fig:rlloss}%
    }
    \\
    \vspace{0.5cm}
    \subfloat[ Average latency]{%
        \includegraphics[width=0.47\columnwidth]{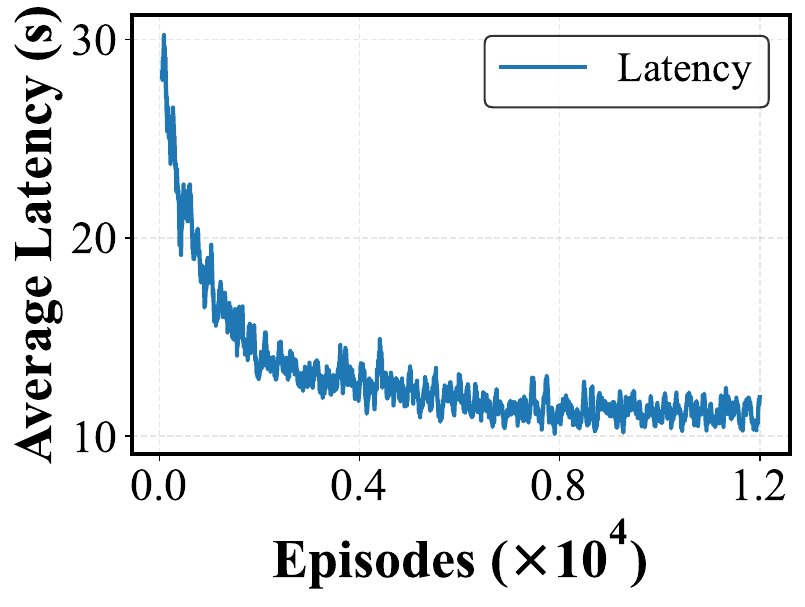}%
        \label{fig:rllatency}%
    }
    \hfil
        \subfloat[Average completion rate ]{%
        \includegraphics[width=0.47\columnwidth]{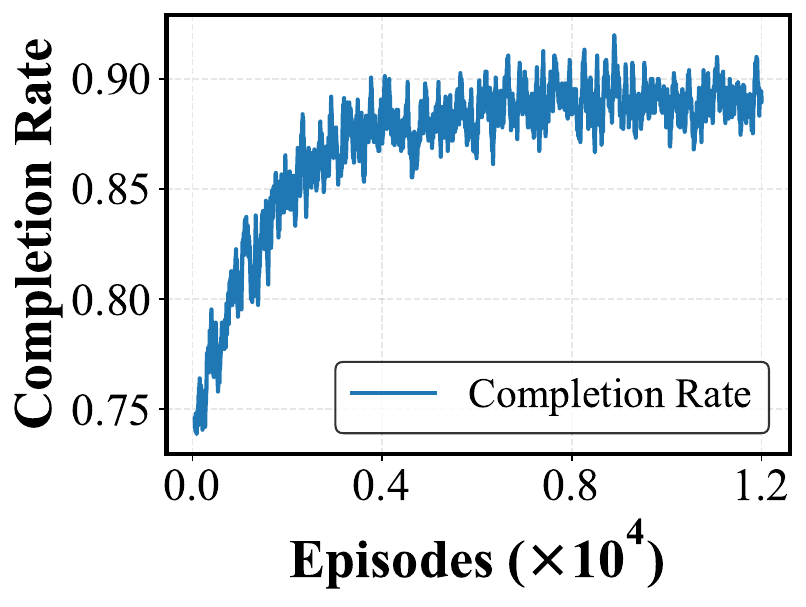}%
        \label{fig:rlcr}%
    }
    \caption{Training convergence of the proposed QLMIO Framework.}
    \label{fig:rl}
\end{figure}

We illustrate the convergence performance of the proposed QLMIO framework in a scenario with 15 servers (cf. Table \ref{tab:system_config}) and 30 users, and the results are displayed in Fig. \ref{fig:rl}. 

As illustrated in Fig. \ref{fig:rlreward}, the average reward of the DRL agent in QLMIO exhibits rapid growth during the initial 4,000 episodes, subsequently stabilizing at approximately 20 by the 10,000 episodes, indicating effective learning. This convergence is further verified by Fig. \ref{fig:rlloss}, where the average training loss plummets to near-zero values, confirming that the DRL agent achieves sufficient convergence during the training process. Concurrently, Fig. \ref{fig:rllatency} reveals that the average task offloading latency undergoes a sharp decline in the first 4,000 episodes, ultimately converging to approximately 10 seconds at the 10,000 episodes. Complementing these results, Fig. \ref{fig:rlcr} shows a steady improvement in the average task completion success rate, which increases from approximately 70\% to 90\% as training progresses. Collectively, these metrics demonstrate that QLMIO can finish the offloading task in a highly balanced fashion, effectively addressing the competing demands of offloading latency and generation quality.

\subsection{Performance Comparison}

\begin{figure*}[!t]
\centering
\begin{subfigure}{0.31\textwidth}
    \includegraphics[width=\linewidth]{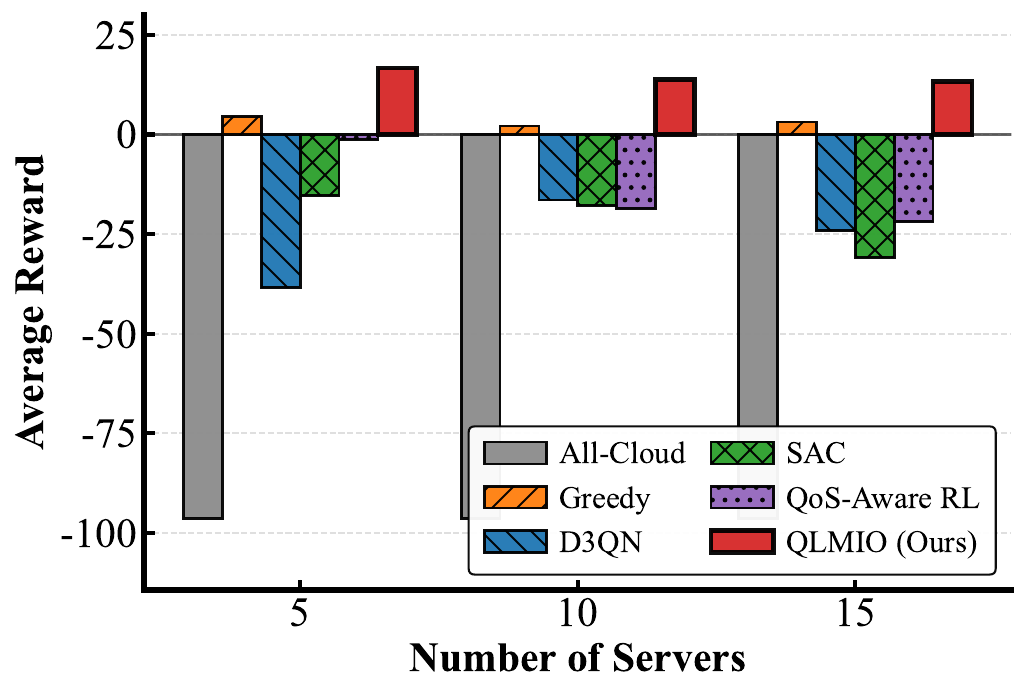}
    \caption{Average system reward vs. server count (30 users)}
    \label{TestSet:sub1}
\end{subfigure}
\hfill
\begin{subfigure}{0.31\textwidth}
    \includegraphics[width=\linewidth]{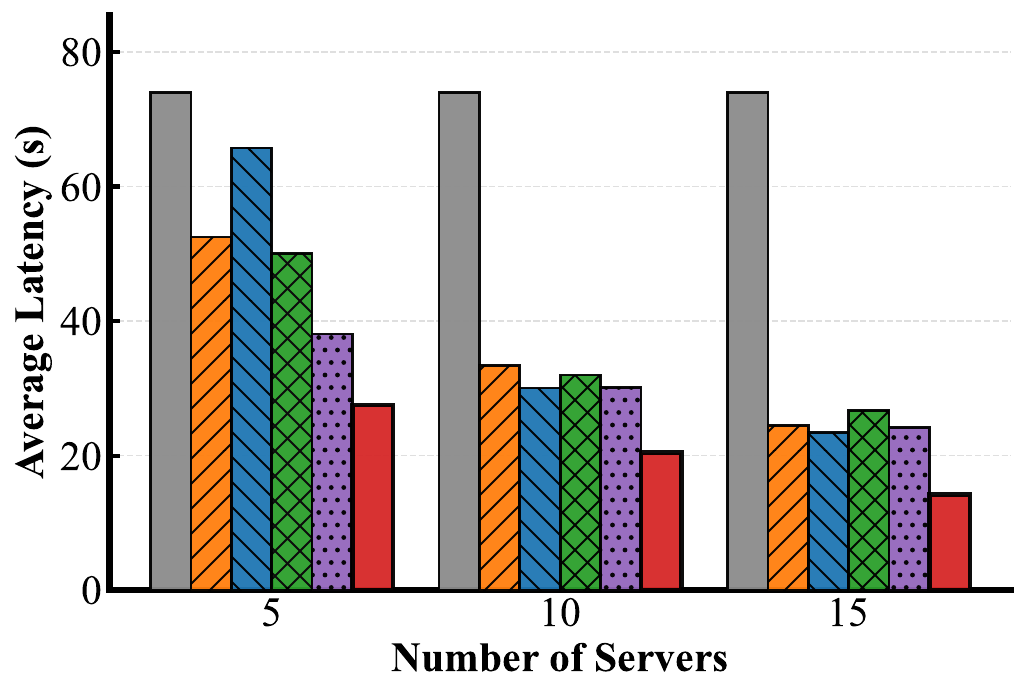}
    \caption{Average offloading latency vs. server count (30 users)}
    \label{TestSet:sub2}
\end{subfigure}
\hfill
\begin{subfigure}{0.31\textwidth}
    \includegraphics[width=\linewidth]{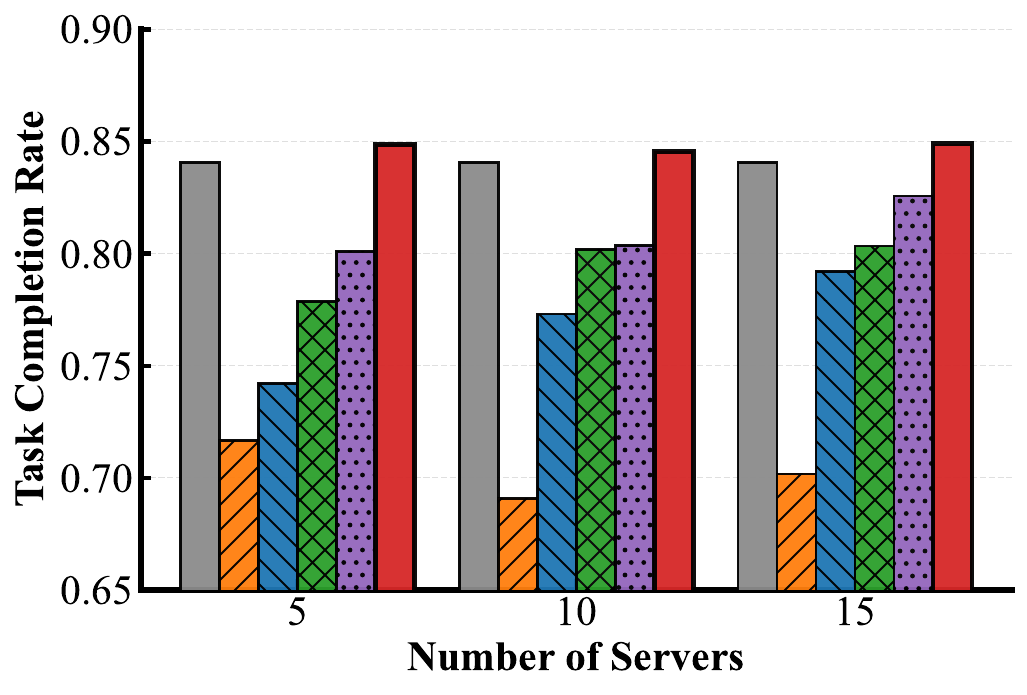}
    \caption{Average task completion rate vs. server count (30 users)}
    \label{TestSet:sub3}
\end{subfigure}

\vspace{0.3cm} 

\begin{subfigure}{0.31\textwidth}
    \includegraphics[width=\linewidth]{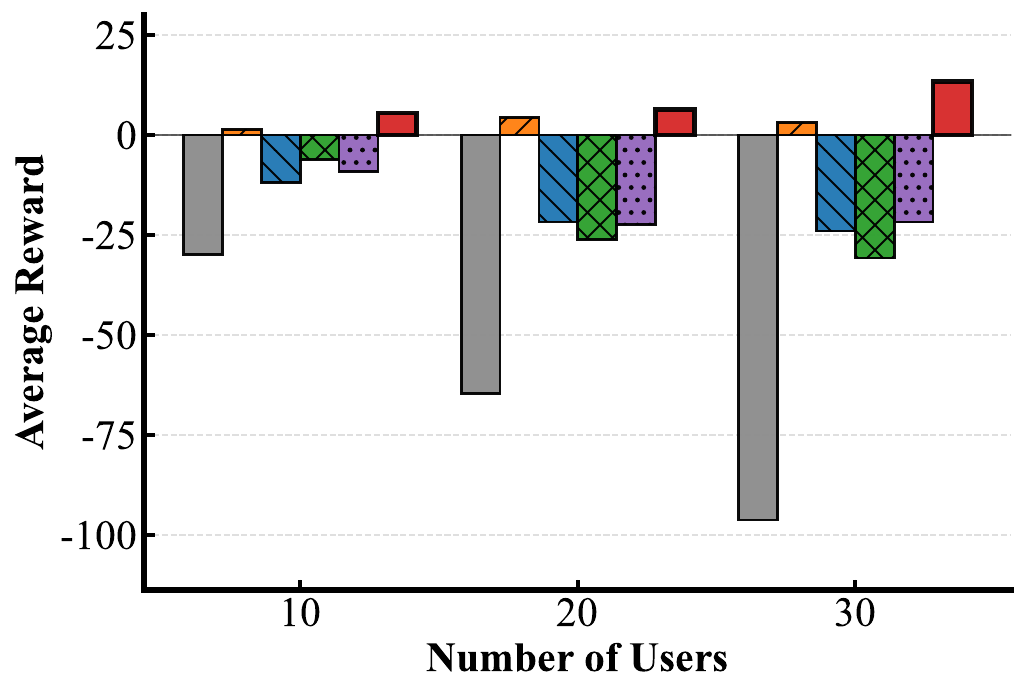}
    \caption{Average system reward vs. user count (15 servers)}
    \label{TestSet:sub4}
\end{subfigure}
\hfill
\begin{subfigure}{0.31\textwidth}
    \includegraphics[width=\linewidth]{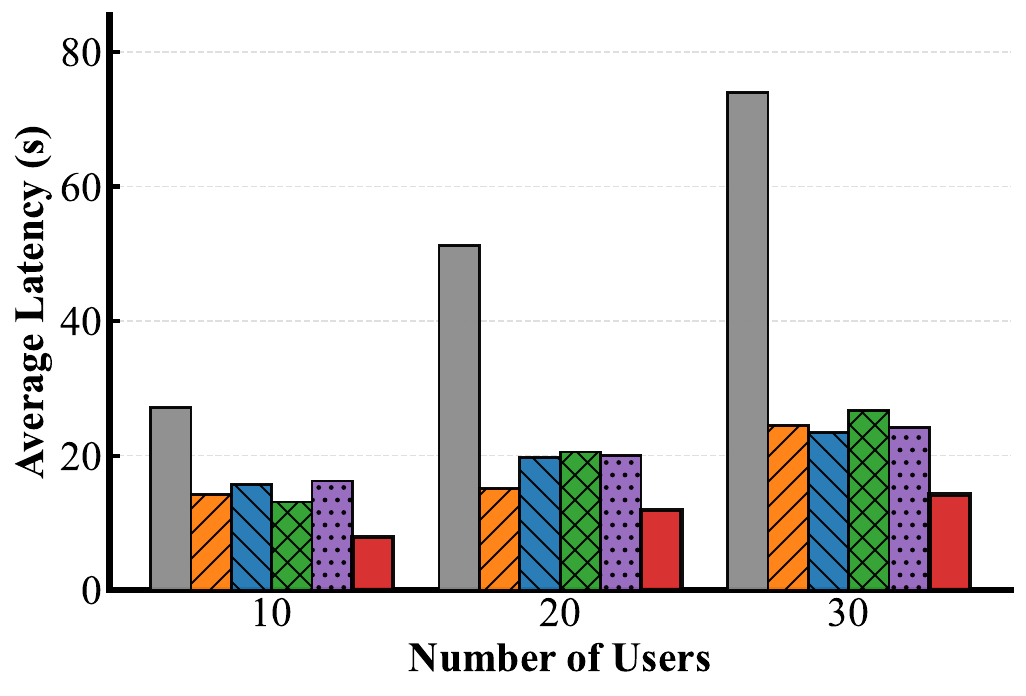}
    \caption{Average offloading latency vs. user count (15 servers)}
    \label{TestSet:sub5}
\end{subfigure}
\hfill
\begin{subfigure}{0.31\textwidth}
    \includegraphics[width=\linewidth]{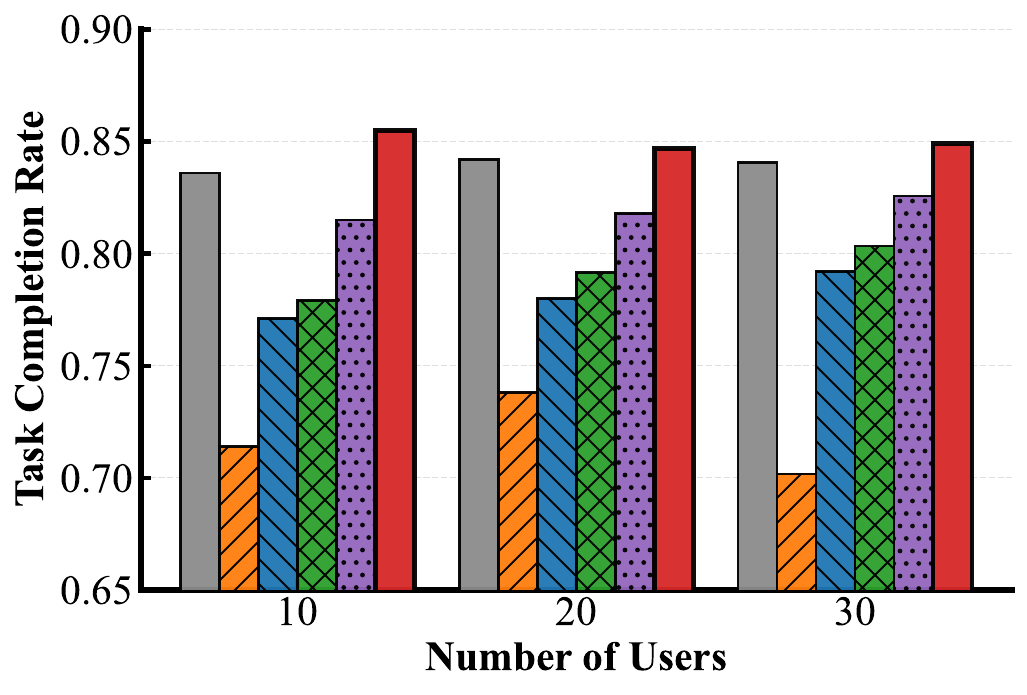}
    \caption{Average task completion rate vs. user count (15 servers)}
    \label{TestSet:sub6}
\end{subfigure}
\caption{Performance of average reward, average offloading latency, and average task completion rate with different server numbers and user numbers.}
\label{TestSet}
\end{figure*}
\label{Sec:Conclusion}

We compare the performance of the proposed QLMIO framework with baselines using the test set. All experiments were conducted through 100 repeated trials, with results aggregated by averaging to provide robust statistical support for our findings, as illustrated in Fig. \ref{TestSet}.

1) Figs. \ref{TestSet:sub1} to \ref{TestSet:sub3} present the performance evaluation of the proposed QLMIO framework under a constant load of 30 users while the number of servers scales from 5 to 15. Fig. \ref{TestSet:sub1} demonstrates that our proposed QLMIO framework achieves the optimal average reward across all server configurations, effectively verifying its superior capability in balancing offloading latency and generation quality. Specifically, from Fig. \ref{TestSet:sub2}, we observe that when the number of edge servers varies between 5 and 15, the proposed QLMIO framework reduces the total task offloading latency (Eq. \ref{offloading}) by 62.83\% to 80.83\%, 27.72\% to 41.43\%, 31.92\% to 58.14\%, 36.06\% to 46.86\%, and 38.79\% to 47.59\%, compared to the All-Cloud, Qos-Aware RL, D3QN, SAC, and Greedy algorithms, respectively.
Simultaneously, Fig. \ref{TestSet:sub3} demonstrates that our QLMIO framework achieves comparable task completion rates to the All-Cloud method and achieves improvements in the average task completion rate ranging from 2.83\%to 5.95\%, 7.20\% to 14.38\%,  5.49\% to 8.99\%, and 18.42\% to 22.44\%, respectively, compared to the remaining QoS-Aware RL, D3QN, SAC, and Greedy baselines. The reason for this performance improvement is that our QLMIO framework utilizes the proposed MILP and MGQP modules to anticipate inference latency and quality before making decisions. This predictive mechanism empowers QLMIO to make judicious offloading decisions that comprehensively account for both inference latency and generation quality. It is worth noting that although QoS-Aware RL also employs models such as linear regression to estimate QoS metrics (e.g., inference latency), it considers only textual information and is therefore not well suited to multimodal inference requests. The lack of multimodal information extraction limits its ability to capture the characteristics of MLLM workloads, leading to performance degradation when handling MLLM inference offloading tasks.

2) Figs. \ref{TestSet:sub4} to \ref{TestSet:sub6} present the performance evaluation of the proposed QLMIO framework under a constant server number of 15 while the number of users scales from 10 to 30. In these scalable user scenarios, we can observe that QLMIO maintains the highest average reward among all evaluated methods from Fig. \ref{TestSet:sub4}, demonstrating its ability to maintain an optimal balance between inference latency and generation quality. From Fig. \ref{TestSet:sub5}, we observe that when the number of users varies between 10 and 30, the proposed QLMIO framework reduces the average task offloading latency by 70.97\% to 80.83\%, 40.57\% to 51.39\%, 39.43\% to 49.81\%, 39.80\% to 46.86\%, and 21.25\% to 44.66\%, compared to the All Cloud, QoS-Aware RL, D3QN, SAC, and Greedy methods, respectively.
Concurrently, Fig. \ref{TestSet:sub6} reveals that our QLMIO achieves a task completion rate of approximately 85\% that of the All-Cloud method, while establishing a robust empirical superiority over the remaining baselines. Specifically, it improves the average task completion rate by 2.83\% to 4.91\%, 7.20\% to 10.89\%, 5.68\% to 9.76\%, and 14.77\% to 21.00\% compared to the QoS-Aware RL, D3QN, SAC, and Greedy methods, respectively. The reason for this phenomenon extends beyond the fact that QLMIO employs the MILP and MGQP modules to perceive inference latency and generation quality. It also derives from the capability of the encompassed DRL architecture to effectively manage the impact caused by the fluctuating number of users within the cloud-edge collaborative MLLM system.

Consequently, traditional computation offloading methods suffer from limited capability in extracting the features of inference latency and generation quality inherent in MLLM inference tasks, often leading to suboptimal or inferior performance. By contrast, our proposed QLMIO framework effectively addresses the MLLM inference offloading in various system configurations by integrating the MILP module, the MGQP module, and the DRL architecture.

\begin{figure*}[!t]
\centering
\begin{subfigure}{0.31\textwidth}
    \includegraphics[width=\linewidth]{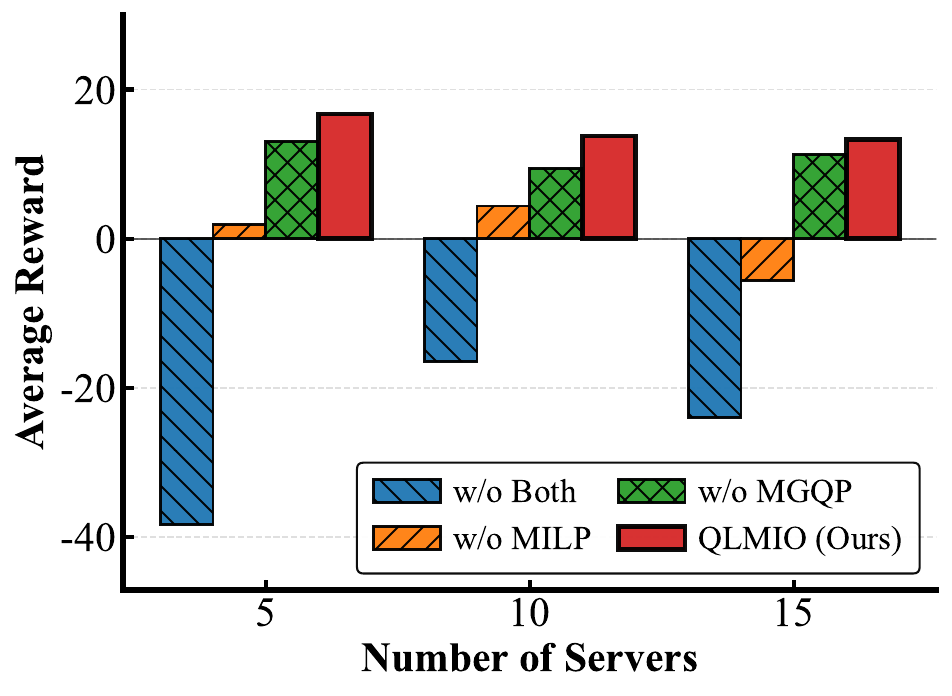}
    \caption{Average system reward vs. server count (30 users)}
    \label{Ablation:sub1}
\end{subfigure}
\hfill
\begin{subfigure}{0.31\textwidth}
    \includegraphics[width=\linewidth]{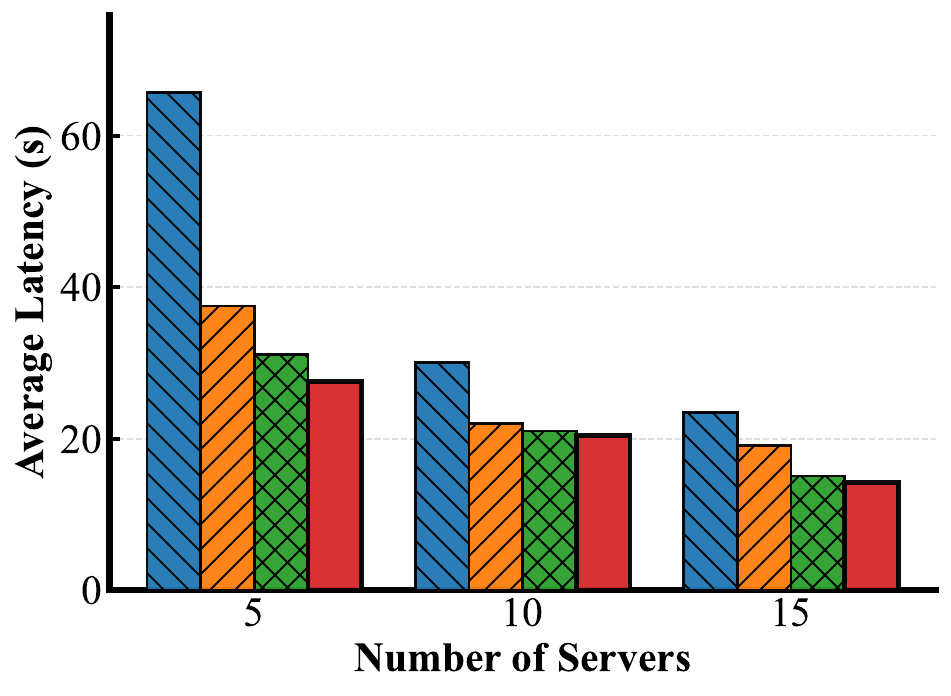}
    \caption{Average offloading latency vs. server count (30 users)}
    \label{Ablation:sub2}
\end{subfigure}
\hfill
\begin{subfigure}{0.31\textwidth}
    \includegraphics[width=\linewidth]{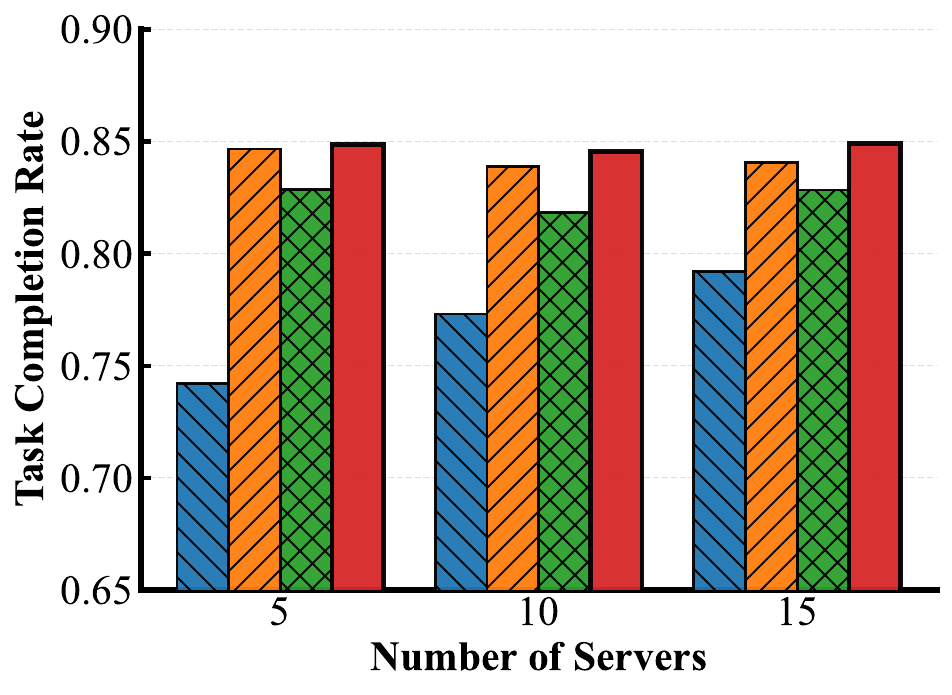}
    \caption{Average task completion rate vs. server count (30 users)}
    \label{Ablation:sub3}
\end{subfigure}

\vspace{0.3cm} 

\begin{subfigure}{0.31\textwidth}
    \includegraphics[width=\linewidth]{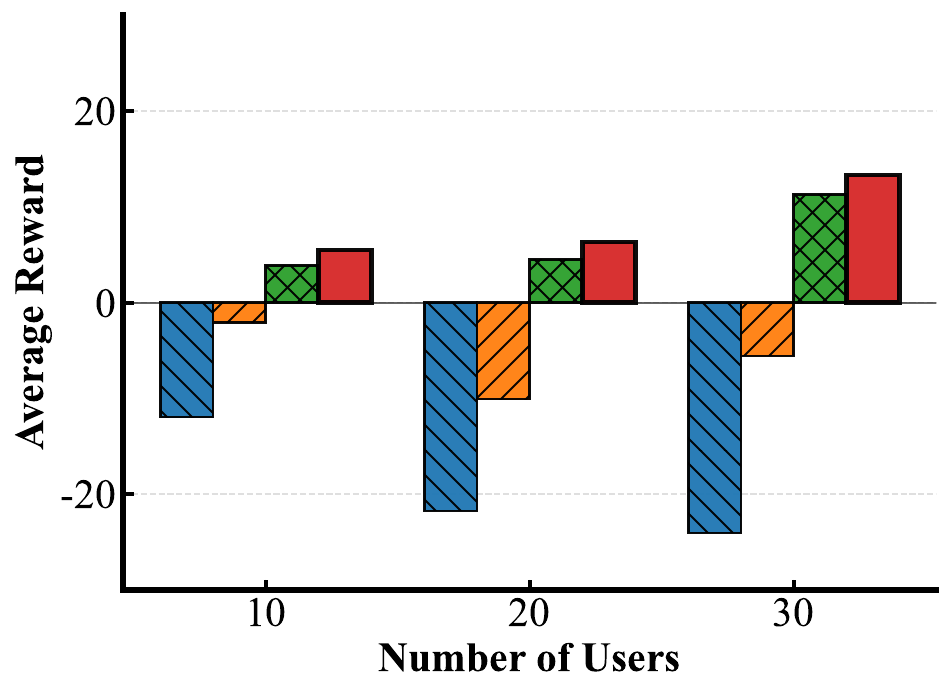}
    \caption{Average system reward vs. user count (30 servers)}
    \label{Ablation:sub4}
\end{subfigure}
\hfill
\begin{subfigure}{0.31\textwidth}
    \includegraphics[width=\linewidth]{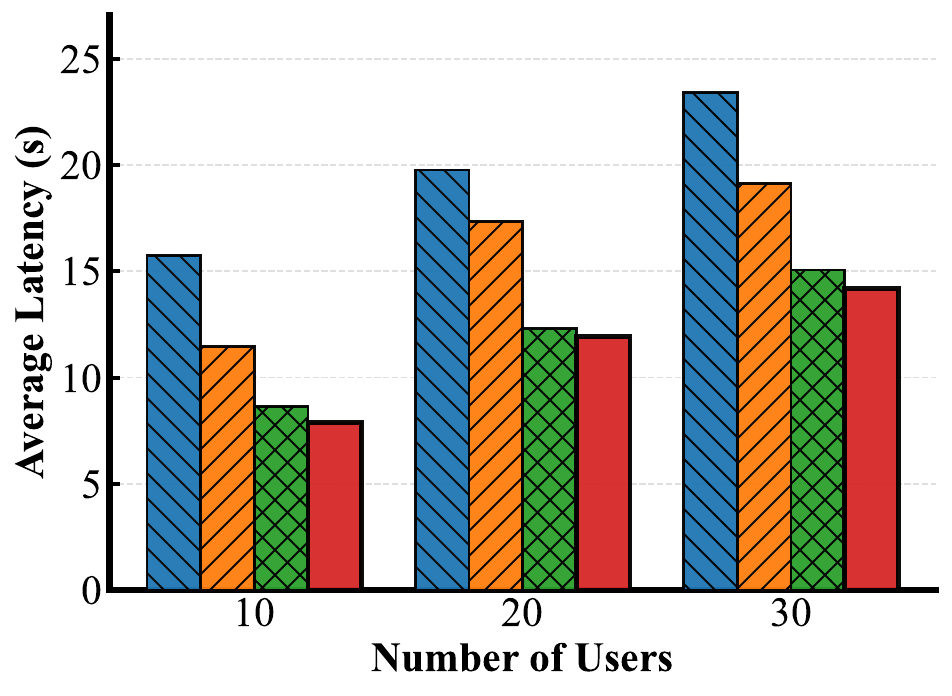}
    \caption{Average offloading latency vs. user count (30 servers)}
    \label{Ablation:sub5}
\end{subfigure}
\hfill
\begin{subfigure}{0.31\textwidth}
    \includegraphics[width=\linewidth]{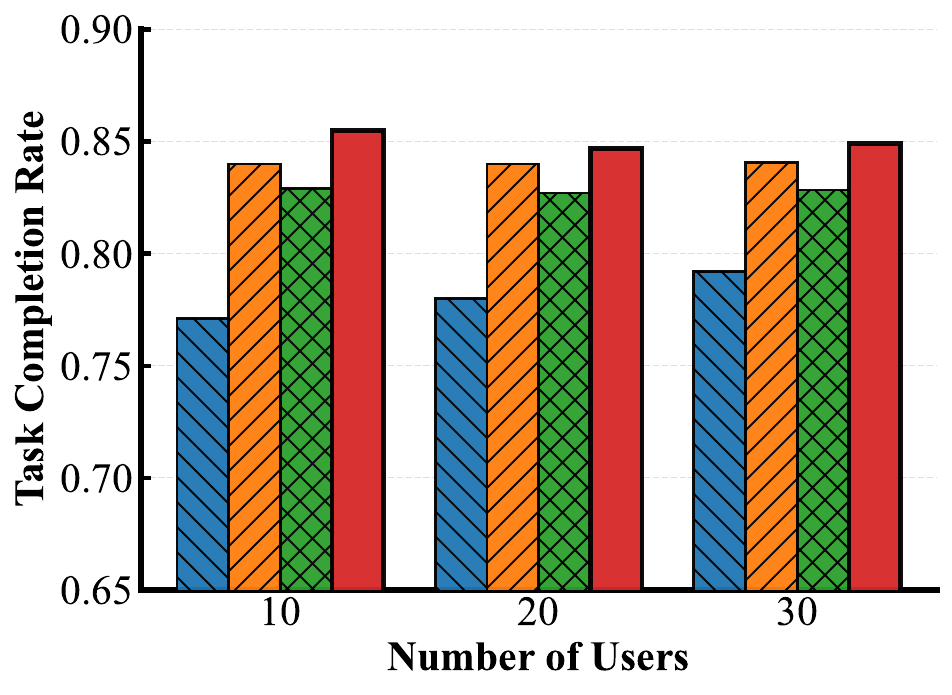}
    \caption{Average task completion rate vs. user count (30 servers)}
    \label{Ablation:sub6}
\end{subfigure}
\caption{Ablation study on MILP and MGQP modules in our QLMIO framework.}
\label{Ablation}
\end{figure*}
\vspace{-0.1in}
\subsection{Ablation Study for QLMIO}

We performed an ablation study to validate the effectiveness of the MILP and MGQP modules in our QLMIO framework. This study evaluated the QLMIO framework with different configurations: 1) without the MILP module, 2) without the MGQP module, and 3) without MILP and MGQP modules, as shown in Fig. \ref{Ablation}.

1) Figs. \ref{Ablation:sub1} and \ref{Ablation:sub4} illustrate the average rewards achieved by the original algorithm and all ablated configurations evaluated on the test set. It is evident that across the four algorithm configurations, the system reward follows the order (from the highest to the lowest): QLMIO, QLMIO without MGQP module, QLMIO without MILP module, and QLMIO without MGQP and MILP modules. This phenomenon substantiates that the proposed QLMIO framework makes offloading decisions that optimally balance inference latency and quality when it jointly integrates the proposed MILP and MGQP modules, thereby ensuring that the system maintains a superior level of QoS.

2) Figs. \ref{Ablation:sub2} and \ref{Ablation:sub5} demonstrate that when the system configuration changes, the proposed QLMIO framework reduces the average offloading latency by 7.08\% to 31.21\%, 2.66\% to 11.50\%, and 31.92\% to 58.14\% compared to the QLMIO without MILP module, QLMIO without MGQP module, and QLMIO without MILP and MGQP modules, respectively. It is observable that the performance improvements achieved by QLMIO regarding offloading latency reduction are significantly less pronounced when compared against the variant lacking the MGQP module than against the other two configurations. The reason for this phenomenon is that the QLMIO without MGQP module completely preserves the proposed MILP component to make offloading decisions highly aware of inference latency. As a result, the QLMIO without MGQP module successfully sustains low latency under varying system configurations, consequently further substantiating the effectiveness of the proposed MILP module.

3) Figs. \ref{Ablation:sub3} and \ref{Ablation:sub6} demonstrate that when the system configuration changes, the proposed QLMIO framework increases the average task completion rate by 0.24\% to 1.79\%, 2.24\% to 3.34\%, and 7.20\% to 14.38\% compared to the QLMIO without MILP module, QLMIO without MGQP module, and QLMIO without MILP and MGQP modules, respectively. We note that the QLMIO without the MILP module achieves nearly identical average task completion rates when compared to the original QLMIO framework. The underlying reason is that this ablated version still utilizes the proposed MGQP module to predict generation correctness probabilities for each action before executing an offloading decision. This proactive prediction ensures that the system maintains high completion rates with changing configurations, which strongly demonstrates the validity of the MGQP module.

Consequently, these results robustly validate that both the proposed MILP and the MGQP modules enhance the performance of our QLMIO framework in the MLLM inference offloading problem.

\section{Conclusion and Future Work}

In this paper, we investigated the MLLM inference offloading problem in cloud-edge continuum, and developed a novel framework QLMIO that successfully provides users with low-latency and high-quality MLLM inference capabilities. The developed QLMIO framework addresses several intractable challenges in this problem, such as the joint prediction and optimization for inference latency and generation quality. Specifically, QLMIO utilizes the proposed MILP and MGQP modules to predict inference latency and task success rates before making offloading decisions, thereby generating offloading decisions that optimally balance inference latency and generation quality. We constructed a real-world cloud-edge collaborative MLLM system, which successfully addressed the issue of the absence of any publicly available dataset for the MLLM inference offloading problem. Based on this system, we collected and created the MIOBench dataset that was leveraged to comprehensively train and evaluate our framework. Extensive experimental results substantiate the effectiveness of the proposed MILP and MGQP modules, while demonstrating the remarkable superiority of the QLMIO framework in tackling the MLLM inference offloading problem.

Motivated by the continuous emergence of various MLLMs, on the one hand, our future work intends to enrich the MIOBench with comprehensive samples encompassing more varied task categories and complex server configurations. On the other hand, we plan to explore innovative inference routing mechanisms and resource allocation approaches to enhance the service quality for advanced MLLM applications in cloud-edge continuum.
\label{Sec:Conclusion}
\bibliographystyle{IEEEtran} 
\bibliography{ref} 

@INPROCEEDINGS{mmbench,
author = {Liu, Yuan and Duan, Haodong and Zhang, Yuanhan and Li, Bo and Zhang, Songyang and Zhao, Wangbo and Yuan, Yike and Wang, Jiaqi and He, Conghui and Liu, Ziwei and Chen, Kai and Lin, Dahua},
title = {MMBench: Is Your Multi-modal Model an All-around Player?},
year = {2024},
booktitle = {2024 European Conference on Computer Vision},
pages = {216–233},
}

@INPROCEEDINGS{EDGELLM_ICWS2024,
  author={Cai, Fenglong and Yuan, Dong and Yang, Zhe and Cui, Lizhen},
  booktitle={2024 IEEE International Conference on Web Services}, 
  title={Edge-LLM: A Collaborative Framework for Large Language Model Serving in Edge Computing}, 
  year={2024},
  volume={},
  number={},
  pages={799-809}}

@INPROCEEDINGS{EDGELLM_ICWS2025,
  author={Jin, Hongpeng and Wu, Yanzhao},
  booktitle={2025 IEEE International Conference on Web Services}, 
  title={CE-CoLLM: Efficient and Adaptive Large Language Models Through Cloud-Edge Collaboration}, 
  year={2025},
  volume={},
  number={},
  pages={316-323},
  doi={10.1109/ICWS67624.2025.00046}}

@ARTICLE{EDGELLM_IOTJ2025,
  author={Zhang, Mingjin and Shen, Xiaoming and Cao, Jiannong and Cui, Zeyang and Jiang, Shan},
  journal={IEEE Internet of Things Journal}, 
  title={EdgeShard: Efficient LLM Inference via Collaborative Edge Computing}, 
  year={2025},
  volume={12},
  number={10},
  pages={13119-13131}}

@ARTICLE{EDGELLM_TWC2025,
  author={Zhang, Xinyuan and Nie, Jiangtian and Huang, Yudong and Xie, Gaochang and Xiong, Zehui and Liu, Jiang and Niyato, Dusit and Shen, Xuemin},
  journal={IEEE Transactions on Wireless Communications}, 
  title={Beyond the Cloud: Edge Inference for Generative Large Language Models in Wireless Networks}, 
  year={2025},
  volume={24},
  number={1},
  pages={643-658}}

@INPROCEEDINGS{COT_ICLR2025,
  title={When More is Less: Understanding Chain-of-Thought Length in LLMs},
  author={Yuyang, Wu and Yifei, Wang and Tianqi, Du and Stefanie, Jegelka and Yisen, Wang},
  booktitle={ ICLR 2025 Workshop on Reasoning and Planning for LLMs},
  year={2025},
  pages={1-21}
}

@ARTICLE{EDGELLM_TNSE2026,
  author={Zheng, Xixi and Li, You and Zheng, Baokun and Zhang, Chuan and Zhu, Liehuang},
  journal={IEEE Transactions on Network Science and Engineering}, 
  title={EdgeNetLLM: Cloud–Edge Collaborative Adaptation of Large Language Models for Mobile Networking}, 
  year={2026},
  volume={13},
  number={1},
  pages={3928-3943}}

@ARTICLE{Offloading0_TMC2025,
  author={Wu, Haixing and Zheng, Jiameng and Jin, Shunfu},
  journal={IEEE Transactions on Mobile Computing}, 
  title={Adaptive Computation Offloading Scheme Based on a Collaborative Architecture With Heterogeneous MEC Nodes: A DRL Approach}, 
  year={2025},
  volume={24},
  number={11},
  pages={12692-12710},
  doi={10.1109/TMC.2025.3586623}}

@ARTICLE{Offloading1_TMC2025,
  author={Chen, Ying and Yang, Yaozong and Hu, Jintao and Wu, Yuan and Huang, Jiwei},
  journal={IEEE Transactions on Mobile Computing}, 
  title={A Game-Theoretical Approach for Distributed Computation Offloading in LEO Satellite-Terrestrial Edge Computing Systems}, 
  year={2025},
  volume={24},
  number={5},
  pages={4389-4402},
  doi={10.1109/TMC.2025.3526200}}

@INPROCEEDINGS{llmoffload_VTC2024,
  author={Wang, Yitong and Liu, Chang and Zhao, Jun},
  booktitle={2024 IEEE 99th Vehicular Technology Conference}, 
  title={Offloading and Quality Control for AI Generated Content Services in 6G Mobile Edge Computing Networks}, 
  year={2024},
  volume={},
  number={},
  pages={1-7},
  doi={10.1109/VTC2024-Spring62846.2024.10683477}}

@ARTICLE{llmoffload_TMC2024,
  author={He, Ying and Fang, Jingcheng and Yu, F. Richard and Leung, Victor C.},
  journal={IEEE Transactions on Mobile Computing}, 
  title={Large Language Models (LLMs) Inference Offloading and Resource Allocation in Cloud-Edge Computing: An Active Inference Approach}, 
  year={2024},
  volume={23},
  number={12},
  pages={11253-11264},
  doi={10.1109/TMC.2024.3415661}}

@ARTICLE{llmoffload0_TMC2025,
  author={Yang, Jin and Wu, Qiong and Feng, Zhiying and Zhou, Zhi and Guo, Deke and Chen, Xu},
  journal={IEEE Transactions on Mobile Computing}, 
  title={Quality-of-Service Aware LLM Routing for Edge Computing With Multiple Experts}, 
  year={2025},
  volume={24},
  number={12},
  pages={13648-13662},
  doi={10.1109/TMC.2025.3590969}}

@ARTICLE{llmoffload1_TMC2026,
  author={Xu, Xinyi and Feng, Gang and Liu, Yijing and Qin, Shuang and Wang, Jian and Wang, Yunxiang},
  journal={IEEE Transactions on Mobile Computing}, 
  title={Joint Inference Offloading and Model Caching for Small and Large Language Model Collaboration}, 
  year={2026},
  volume={25},
  number={2},
  pages={2691-2706},
  doi={10.1109/TMC.2025.3608303}}

@ARTICLE{llmoffload_iotj2025,
  author={Huang, Hualong and Du, Yongkang and Zhan, Wenhan and Duan, Hancong and Peng, Kai and Cheng, Yamin and Ye, Yalan and Zhao, Zitian},
  journal={IEEE Internet of Things Journal}, 
  title={Dynamic Model Deployment, Batch Scheduling, and Resource Allocation in MLLM-Enabled Edge–Cloud Networks: A Multiagent Two-Timescale DRL Approach}, 
  year={2025},
  volume={12},
  number={23},
  pages={50818-50835},
  doi={10.1109/JIOT.2025.3611003}}

@ARTICLE{Multiobject_TVT2025,
  author={Xiao, Xiang-Jie and Wang, Yong and Huang, Pei-Qiu and Wang, Kezhi},
  journal={IEEE Transactions on Vehicular Technology}, 
  title={Neural Combinatorial Optimization for Multiobjective Task Offloading in Mobile Edge Computing}, 
  year={2025},
  volume={74},
  number={7},
  pages={10869-10880},
  doi={10.1109/TVT.2025.3546914}}

@ARTICLE{Multiobject1_TVT2025,
  author={Chen, Shuaijie and Li, Wenfeng and Sun, Jingtao and Pace, Pasquale and He, Lijun and Fortino, Giancarlo},
  journal={IEEE Transactions on Vehicular Technology}, 
  title={An Efficient Collaborative Task Offloading Approach Based on Multi-Objective Algorithm in MEC-Assisted Vehicular Networks}, 
  year={2025},
  volume={74},
  number={7},
  pages={11249-11263},
  doi={10.1109/TVT.2025.3543412}}

@INPROCEEDINGS{Sanh2019DistilBERTAD,
  title={DistilBERT, a distilled version of BERT: smaller, faster, cheaper and lighter},
  author={Victor Sanh and Lysandre Debut and Julien Chaumond and Thomas Wolf},
  booktitle={NeurIPS 2019 Workshop on Energy Efficient Machine Learning and Cognitive Computing},
  year={2019},
  pages={1-5}
}

@INPROCEEDINGS{VIT,
  author= {Alexey Dosovitskiy and
                  Lucas Beyer and
                  Alexander Kolesnikov and
                  Dirk Weissenborn and
                  Xiaohua Zhai and
                  Thomas Unterthiner and
                  Mostafa Dehghani and
                  Matthias Minderer and
                  Georg Heigold and
                  Sylvain Gelly and
                  Jakob Uszkoreit and
                  Neil Houlsby},
  title        = {An Image is Worth 16x16 Words: Transformers for Image Recognition
                  at Scale},
  booktitle    = {9th International Conference on Learning Representations},
  year         = {2021},
  pages = {1-22}
}

@ARTICLE{FocalLoss_TMC2026,
  author={Liu, Daibo and Lu, Hang and Qi, Xiaomeng and Rong, Huigui and Chen, Haowen and Jiang, Hongbo},
  journal={IEEE Transactions on Mobile Computing}, 
  title={Decoding Air Friction Rhythms: Enabling User Identification With Out-Ear Microphones in COTS Earphones}, 
  year={2026},
  volume={25},
  number={4},
  pages={4648-4663},
  doi={10.1109/TMC.2025.3626359}}

@ARTICLE{HuberLoss_TMC2025,
  author={Zhang, Haoyu and Zhang, Dongheng and Song, Ruiyuan and Wu, Zhi and Chen, Jinbo and Fang, Liang and Lu, Zhi and Hu, Yang and Lin, Hui and Chen, Yan},
  journal={IEEE Transactions on Mobile Computing}, 
  title={UMIMO: Universal Unsupervised Learning for Mmwave Radar Sensing With MIMO Array Synthesis}, 
  year={2025},
  volume={24},
  number={8},
  pages={7042-7058},
  doi={10.1109/TMC.2025.3546757}}

@ARTICLE{D3QN_TMC2025,
  author={Yang, Meiyi and Gao, Deyun and Zhang, Weiting and Yang, Dong and Niyato, Dusit and Zhang, Hongke and Leung, Victor C. M.},
  journal={IEEE Transactions on Mobile Computing}, 
  title={Deep Reinforcement Learning-Based Joint Caching and Routing in AI-Driven Networks}, 
  year={2025},
  volume={24},
  number={3},
  pages={1322-1337},
  doi={10.1109/TMC.2024.3481276}}

@ARTICLE{egreedy_TPDS2024,
  author={Chen, Xing and Hu, Shengxi and Yu, Chujia and Chen, Zheyi and Min, Geyong},
  journal={IEEE Transactions on Parallel and Distributed Systems}, 
  title={Real-Time Offloading for Dependent and Parallel Tasks in Cloud-Edge Environments Using Deep Reinforcement Learning}, 
  year={2024},
  volume={35},
  number={3},
  pages={391-404},
  doi={10.1109/TPDS.2023.3349177}}

@ARTICLE{softupdate_TNSE2025,
  author={Zhang, Mengyuan and Fang, Juan and Teng, Ziyi and Liu, Yaqi and Wu, Shen},
  journal={IEEE Transactions on Network and Service Management}, 
  title={Joint DNN Partitioning and Task Offloading Based on Attention Mechanism-Aided Reinforcement Learning}, 
  year={2025},
  volume={22},
  number={3},
  pages={2914-2927},
  doi={10.1109/TNSM.2025.3561739}}

@ARTICLE{D3QN_TPDS2026,
  author={Li, Chunlin and Wang, Jiaqi and Jiang, Kun and Xiong, Cheng and Wan, Shaohua},
  journal={IEEE Transactions on Parallel and Distributed Systems}, 
  title={MEOCI: Model Partitioning and Early-Exit Point Selection Joint Optimization for Collaborative Inference in Vehicular Edge Computing}, 
  year={2026},
  volume={37},
  number={3},
  pages={666-679},
  doi={10.1109/TPDS.2026.3652171}}

@ARTICLE{SAC_TSC2025,
  author={Liu, Jianhua and Wang, Xin and Yu, Shui and Xue, Guangtao and Li, Minglu},
  journal={IEEE Transactions on Services Computing}, 
  title={Trustworthy Multi-Hop Cooperative Task Offloading in Device-Edge-Cloud Computing}, 
  year={2025},
  volume={18},
  number={6},
  pages={4304-4317},
  doi={10.1109/TSC.2025.3615156}}

@ARTICLE{Cloud_TNSE2026,
  author={Chang, Yuan and Luan, Tom H. and Wang, Siran and Wang, Yuntao},
  journal={IEEE Transactions on Network Science and Engineering}, 
  title={SafeRAG: Secure Cloud-Based Retrieval-Augmented Generation for LLM-Empowered Voice Assistants}, 
  year={2026},
  volume={13},
  number={1},
  pages={6211-6224},
  doi={10.1109/TNSE.2026.3654070}}

@ARTICLE{LLM_TPDS2026,
  author={Ye, Shengyuan and Ouyang, Bei and Qian, Tianyi and Zeng, Liekang and Li, Jingyi and Du, Jiangsu and Chu, Xiaowen and Xing, Guoliang and Chen, Xu},
  journal={IEEE Transactions on Parallel and Distributed Systems}, 
  title={Resource-Efficient Personal Large Language Models Fine-Tuning With Collaborative Edge Computing}, 
  year={2026},
  volume={37},
  number={3},
  pages={680-696},
  doi={10.1109/TPDS.2026.3654957}}

@ARTICLE{MLLM_TVT2026,
  author={Huang, Shucheng and Shi, Freda and Sun, Chen and Zhong, Jiaming and Ning, Minghao and Yang, Yufeng and Lu, Yukun and Wang, Hong and Khajepour, Amir},
  journal={IEEE Transactions on Vehicular Technology}, 
  title={DriveSOTIF: Advancing SOTIF Through Multimodal Large Language Models}, 
  year={2026},
  volume={75},
  number={3},
  pages={3642-3655},
  doi={10.1109/TVT.2025.3608811}}

@article{Cloudedgellm_TIT2025,
author = {Husom, Erik Johannes and Goknil, Arda and Astekin, Merve and Shar, Lwin Khin and K\~{A}¥sen, Andre and Sen, Sagar and Mithassel, Benedikt Andreas and Soylu, Ahmet},
title = {Sustainable LLM Inference for Edge AI: Evaluating Quantized LLMs for Energy Efficiency, Output Accuracy, and Inference Latency},
year = {2025},
volume = {6},
number = {4},
journal = {ACM Trans. Internet Things},
pages = {1-35}}

@INPROCEEDINGS{Cloudedgellm_icpads2025,
  author={Ling, Xuan and Mao, Yingchi and Tang, Yu and Zhang, Benteng and Rong, Yi and He, Xiaoming},
  booktitle={2025 IEEE 31th International Conference on Parallel and Distributed Systems}, 
  title={LLM-Driven Cloud-Edge Collaboration for Resilient Multi-UAV Task Planning}, 
  year={2025},
  volume={},
  number={},
  pages={1-8},
  doi={10.1109/ICPADS67057.2025.11323223}}

@ARTICLE{Offloading_TNSE2025,
  author={Zhang, Ruipeng and Feng, Yanxiang and Yang, Yikang and Li, Xiaoling and Li, Hengnian},
  journal={IEEE Transactions on Network Science and Engineering}, 
  title={Learning-Based Deadlock-Free Multi-Objective Task Offloading in Satellite Edge Computing With Data-Dependent Constraints and Limited Buffers}, 
  year={2025},
  volume={12},
  number={1},
  pages={356-368},
  doi={10.1109/TNSE.2024.3496902}}

@ARTICLE{Offloading_TPDS2025,
  author={Xiao, Bohuai and Yu, Chujia and Chen, Xing and Chen, Zheyi and Min, Geyong},
  journal={IEEE Transactions on Parallel and Distributed Systems}, 
  title={Multi-Agent Collaboration for Workflow Task Offloading in End-Edge-Cloud Environments Using Deep Reinforcement Learning}, 
  year={2025},
  volume={36},
  number={11},
  pages={2281-2296},
  doi={10.1109/TPDS.2025.3606001}}

@ARTICLE{MSE_TSC2025,
  author={Tummala, Veera Manikantha Rayudu and Kumar, Prince and Tejaswani, Aithi and Roy, Arijit},
  journal={IEEE Transactions on Services Computing}, 
  title={CuPric: Customer Segment-Based Pricing in Mobile Sensors-as-a-Service}, 
  year={2025},
  volume={18},
  number={6},
  pages={3623-3634},
  doi={10.1109/TSC.2025.3615838}}

@ARTICLE{MAE_TPDS2025,
  author={Liao, Haoyu and Liu, Tong and Guo, Jianmei and Huang, Bo and Yang, Dingyu and Ding, Jonathan},
  journal={IEEE Transactions on Parallel and Distributed Systems}, 
  title={Retrospecting Available CPU Resources: SMT-Aware Scheduling to Prevent SLA Violations in Data Centers}, 
  year={2025},
  volume={36},
  number={1},
  pages={67-83},
  doi={10.1109/TPDS.2024.3494879}}

@misc{yang2026msaoadaptivemodalitysparsityaware,
      title={MSAO: Adaptive Modality Sparsity-Aware Offloading with Edge-Cloud Collaboration for Efficient Multimodal LLM Inference}, 
      author={Zheming Yang and Qi Guo and Jun Wan and Jiarui Ruan and Yunqing Hu and Chang Zhao and Xiangyang Li},
      year={2026},
      eprint={2604.02945},
      archivePrefix={arXiv},
      primaryClass={cs.DC},
      url={https://arxiv.org/abs/2604.02945}, 
        pages={1-10},
}

\section{Biography Section}

\begin{IEEEbiography}[{\includegraphics[width=1in,height=1.25in,clip,keepaspectratio]{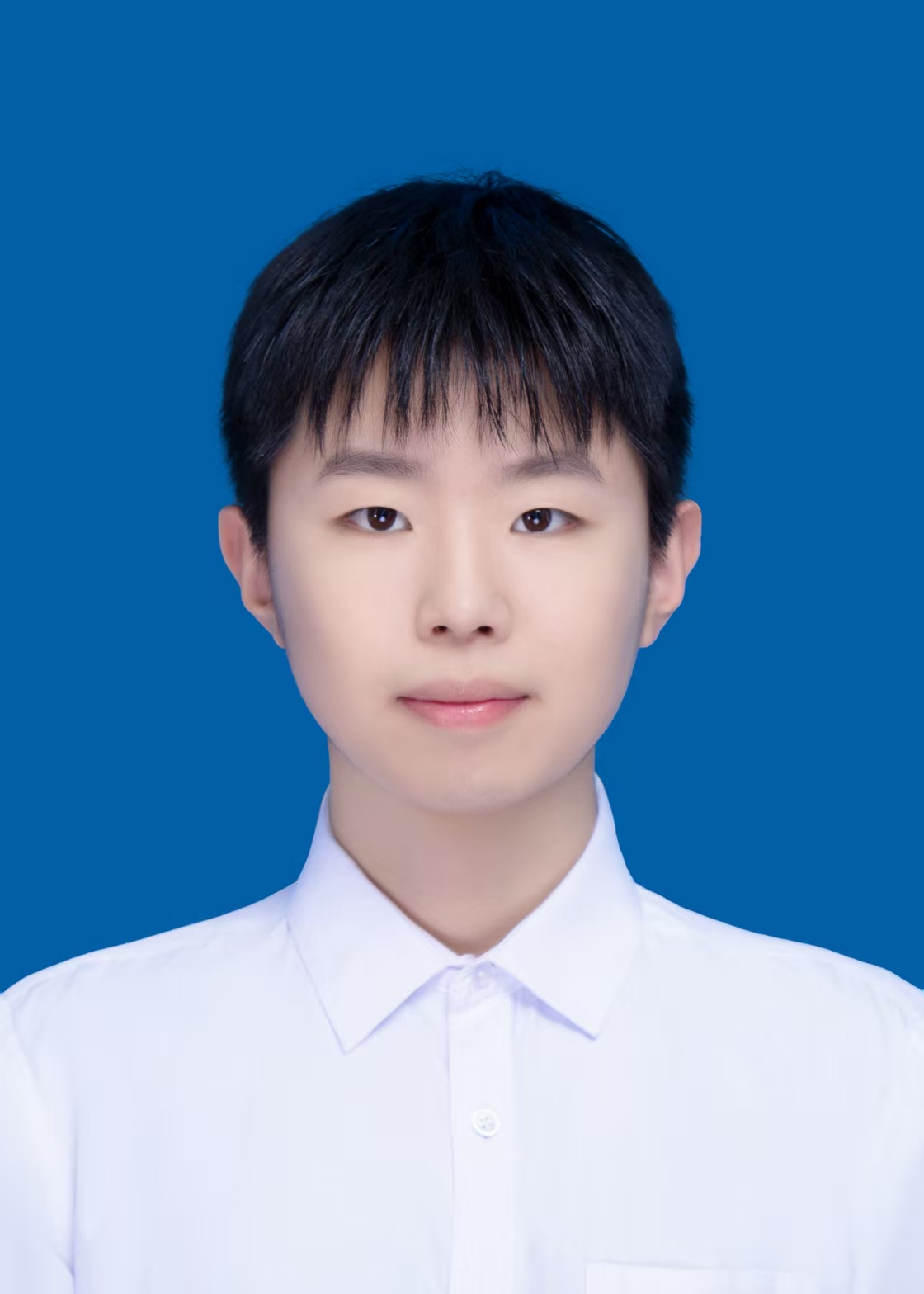}}]{Zhongxiao Wang}
 is a postgraduate student at School of Computer Science and Technology, Xidian University. He is expected to obtain his master degree in 2028. His research interests include edge computing and LLM inference. He obtained the Best Demonstration Award at ICSOC 2025.
\end{IEEEbiography}

\begin{IEEEbiography}[{\includegraphics[width=1in,height=1.25in,clip,keepaspectratio]{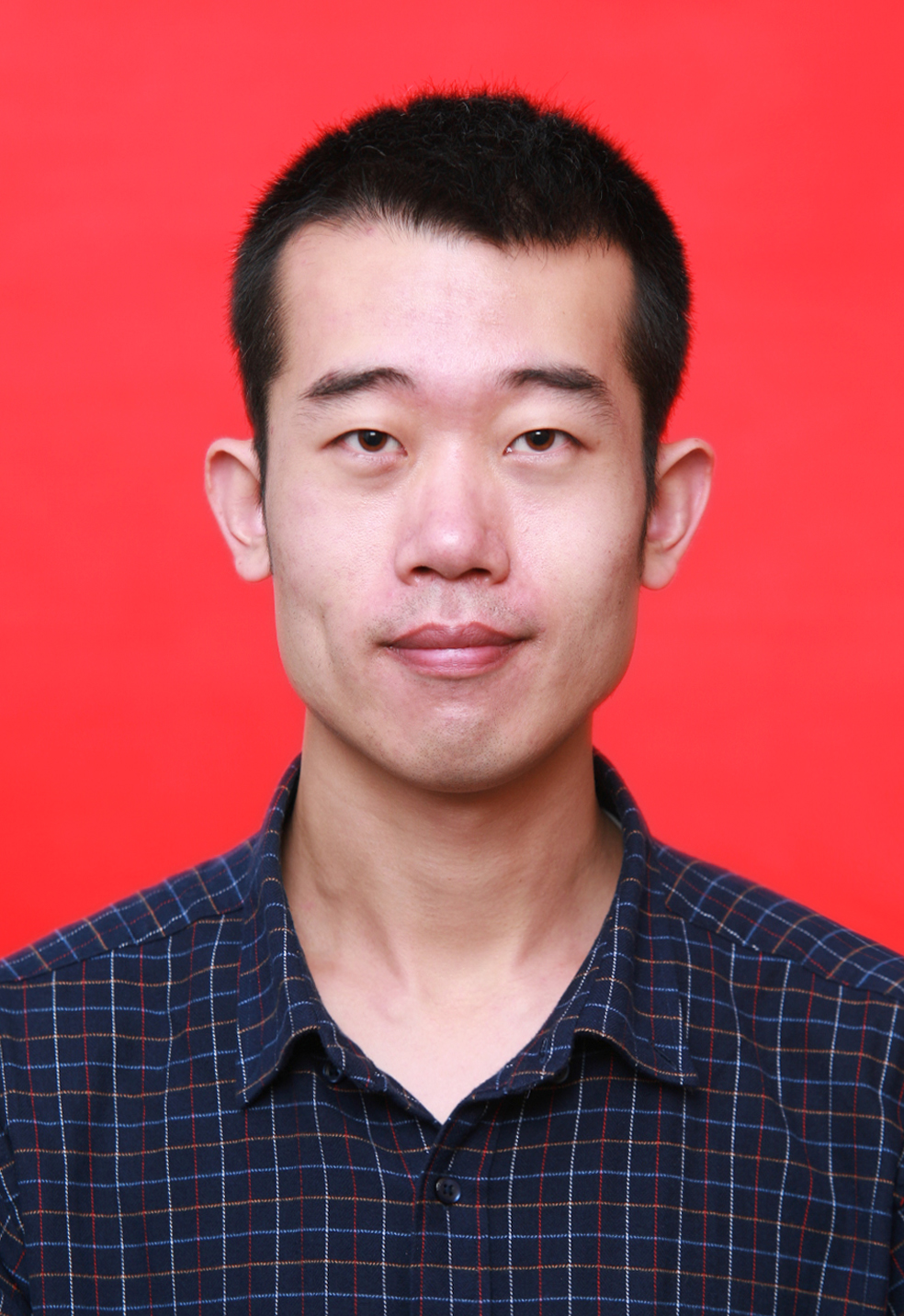}}]{Yueshen Xu} (Member, IEEE)
is an associate professor in School of Computer Science and Technology, Xidian University. He received his Ph.D. degree from Zhejiang University, and was a visiting scholar in University of Illinois at Chicago. His research focuses on edge computing, mobile computing, and LLM. He has published more than 80 papers in prestigious conference and journals such as IEEE TSC, IEEE TOC, IEEE TITS, IEEE TII, IEEE IoT-J, IEEE TVT, IEEE TNSE, IEEE TNSM, WWW, ICSE, and ICSOC. He has won the Best Paper Runner-up Award and the Best Demonstration Award in ICSOC 2025. He has several ESI highly-cited papers and more than 3500 citations. He was selected as one of the ``World's Top 2\% Scientists in 2023'' by the research team of Stanford University.
\end{IEEEbiography}

\begin{IEEEbiography}[{\includegraphics[width=1in,height=1.25in, clip,keepaspectratio]{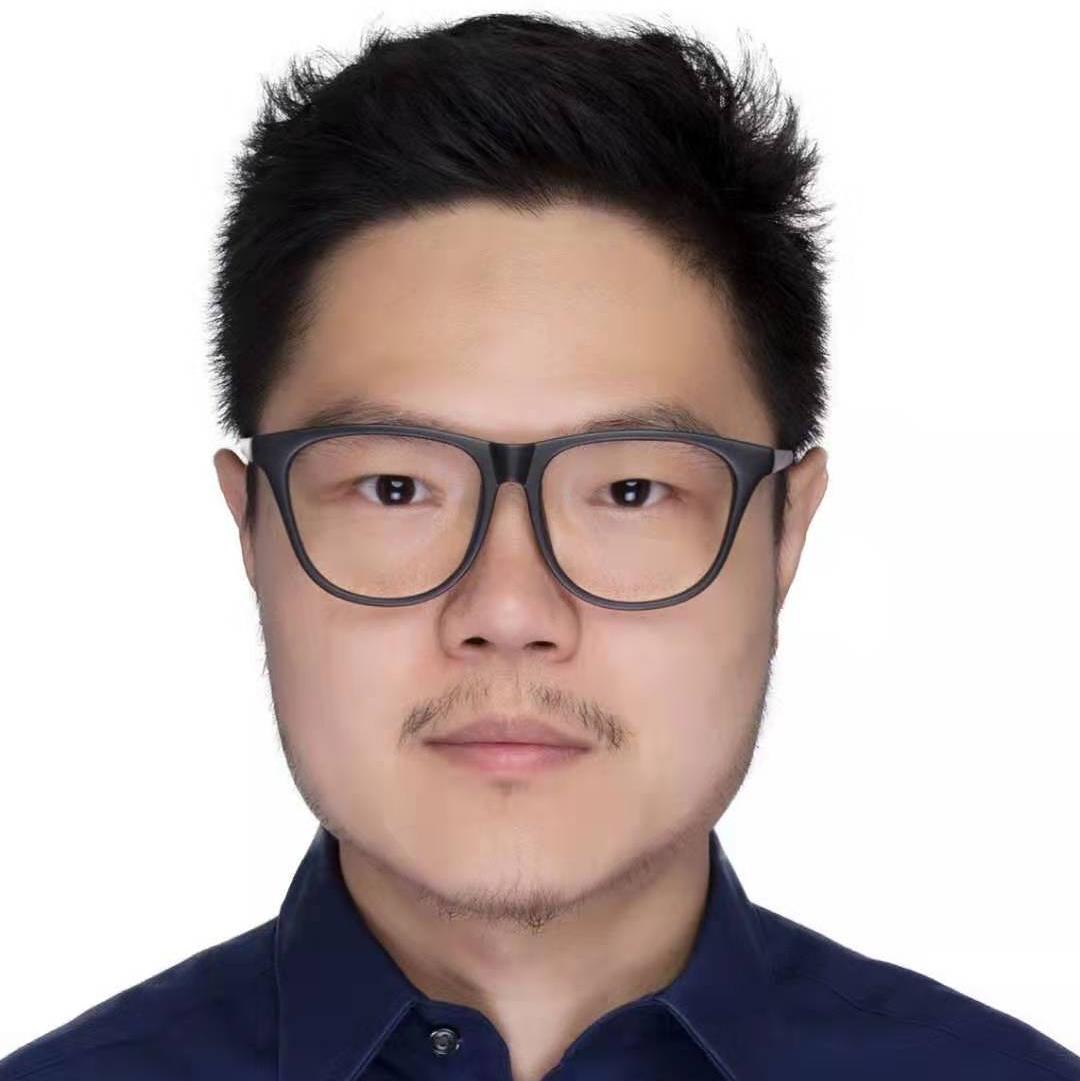}}]{Wei Xi} (Member, IEEE) received the PhD degree in Computer Science from Xi’an Jiaotong University in 2014. He is currently a professor with the School of Computer Science and Technology, Xi’an Jiaotong University. His  research interests include the Internet of Things, mobile computing, and network security. He has published more than 70 papers in leading journals and conferences such as IEEE TMC, IEEE TKDE, IEEE TIFS, IEEE TNNLS, IEEE IoT-J, MobiCom, Ubicomp, INFOCOM, AAAI, ICDCS, and SenSys. 
\end{IEEEbiography}

\begin{IEEEbiography}[{\includegraphics[width=1in,height=1.25in,clip,keepaspectratio]{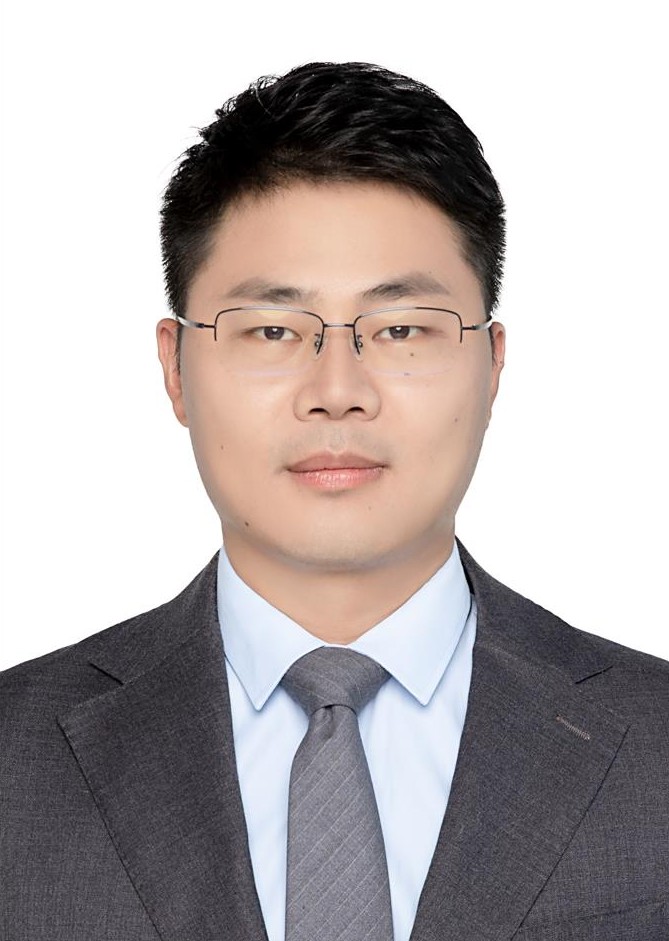}}]{Xinkui Zhao}
is a ZJU 100-Young professor at Zhejiang University. He received his PhD degree from the College of Computer Science and Technology at Zhejiang University, and then worked as the director of container services at Huawei Cloud. His research interests include cloud-native architecture and intelligent software systems. He has led several enterprise cloud-native platforms and authored or coauthored more than 60 papers published on leading journals and conferences such as IEEE TOC, IEEE TSC, IEEE TDCS, IEEE TITS, IEEE TVCG, AAAI, CVPR, WWW, DAC, and ICDE. 
\end{IEEEbiography}

\begin{IEEEbiography}[{\includegraphics[width=1in,height=1.25in,clip,keepaspectratio]{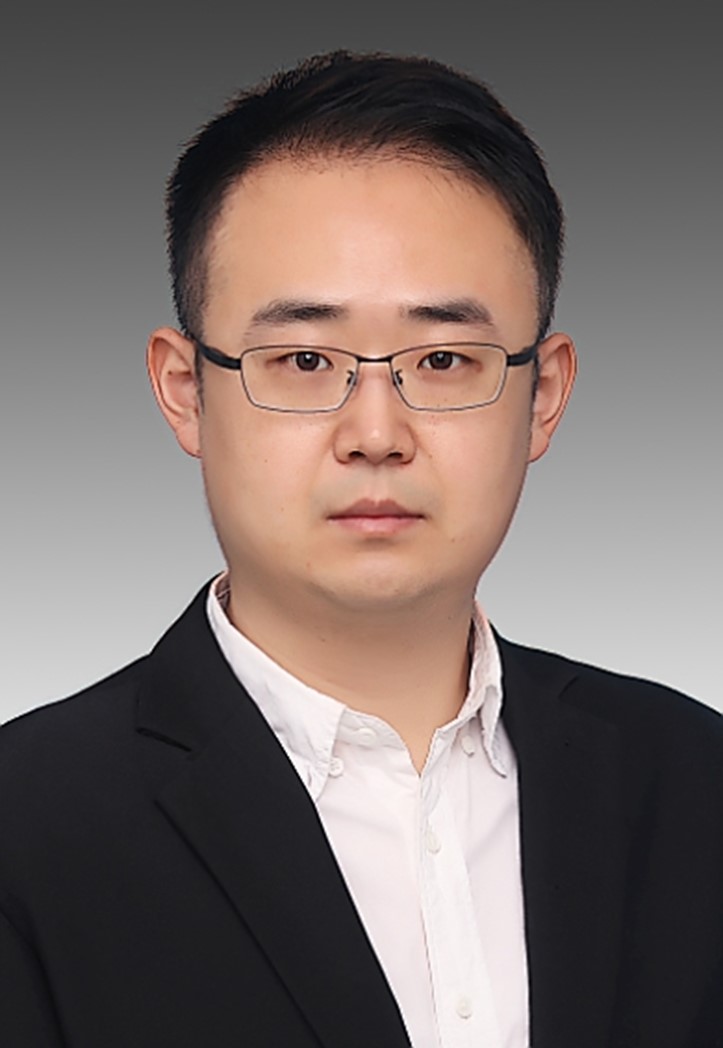}}]{Tom H. Luan} (Fellow, IEEE) received the Ph.D. degree from the University of Waterloo, Canada, in 2012. He is currently a Professor with the School of Cyber Science and Engineering, Xi’an Jiaotong University, China. He has authored or co-authored more than 100 articles in prestigious journals and conferences. He awarded one U.S. patent. His research focuses on wireless networks, cloud computing, and edge computing. He served as a TPC Member for IEEE Globecom and ICC.
\end{IEEEbiography}

\begin{IEEEbiography}[{\includegraphics[width=1in,height=1.25in,clip,keepaspectratio]{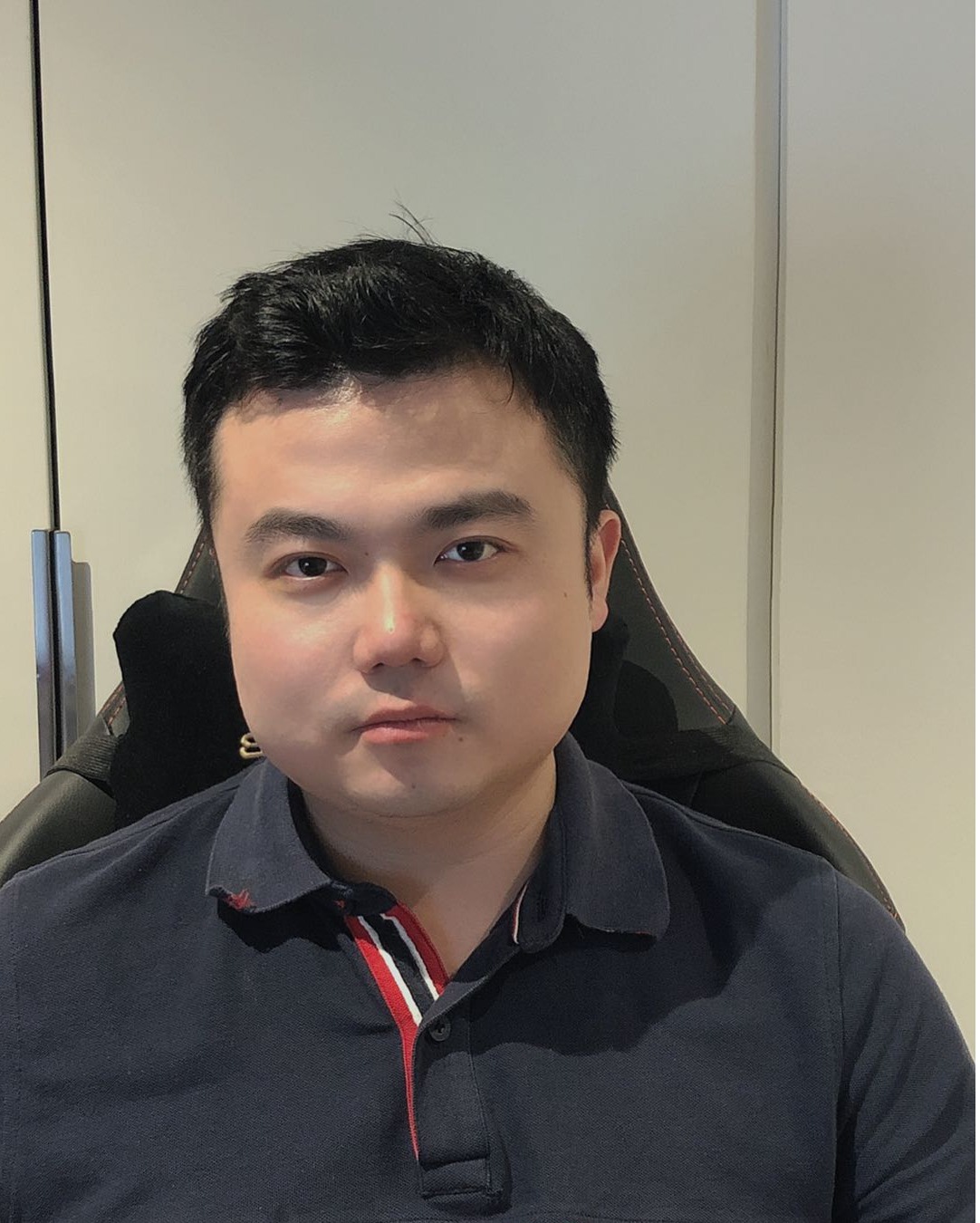}}]{Wei Shao}  (Member, IEEE) is a Lecturer at the University of New South Wales (UNSW). Previously, he served as a Postdoctoral Researcher at UC Davis and Arizona State University. He completed his Ph.D. degree in Computer Science in 2018 from the RMIT University, Australia. His research interests encompass Cybersecurity, Graph Neural Networks, Spatio-temporal Data Mining, Reinforcement Learning Applications, and the Internet of Things (IoT). Wei has authored over 60 research papers published in leading conferences and journals. His contributions to academia have been recognized with accolades such as the Distinguished Paper Award at UbiComp and the Best Reviewer Award at KDD. Additionally, he received CSIRO’s Early Career in Science Award for his outstanding scientific contributions to Australian research.
\end{IEEEbiography}


\begin{IEEEbiography}[{\includegraphics[width=1in,height=1.25in,clip,keepaspectratio]{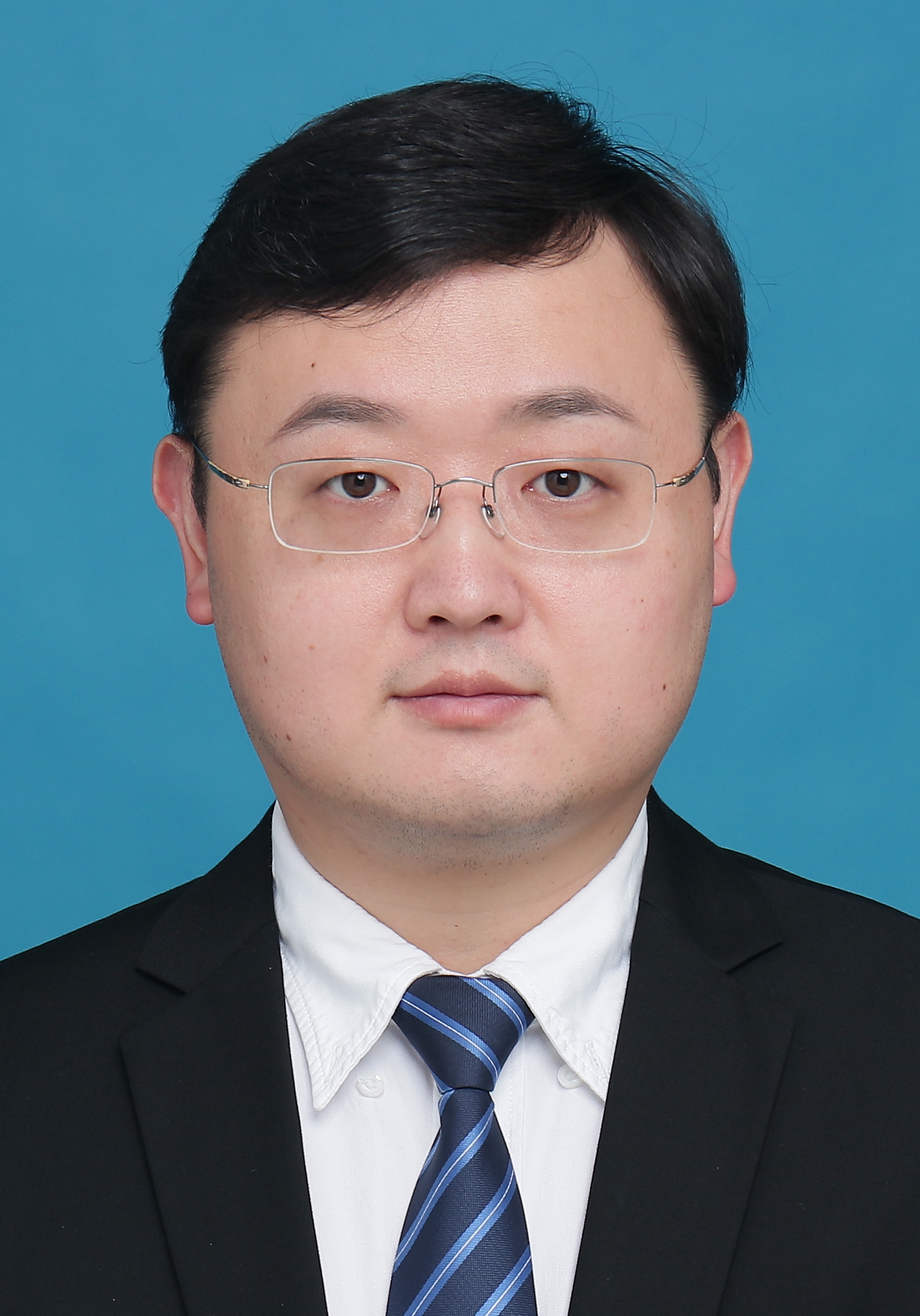}}]{Rui Li} (Member, IEEE) obtained the bachelor degree from Xidian University, at the Department of Applied Mathematics. He obtained his Ph.D. degree from Xi'an Jiaotong University, at the Department of Computer Science and Technology in 2014. He is currently a professor in School of Computer Science and Technology, Xidian University. His interested topics include mobile computing, edge computing, and LLM. He is a member of academic organizations, including IEEE, CCF, and ACM. He has near 60 papers that have been published in related journals and conferences.
\end{IEEEbiography}

\vfill
\end{document}